\documentclass[12pt]{article}

\usepackage[english]{babel}
\usepackage[T1]{fontenc}
\usepackage[utf8]{inputenc}
\usepackage{amssymb,amsmath,array,hhline,bm,curves}
\usepackage{latexsym}
\usepackage{graphicx}
\usepackage[final]{pdfpages}
\usepackage{verbatim}
\usepackage{enumerate,bbm}

\title{Signatures for $J$-hermitians and $J$-unitaries \\ on Krein spaces with Real structures}

\author{Hermann Schulz-Baldes$^{1}$, Carlos Villegas-Blas$^2$
\\
\\
{\small $^1$ Department Mathematik, Friedrich-Alexander-Universit\"at Erlangen-N\"urnberg, Germany}
\\
{\small $^2$ Instituto de Matematicas, Cuernavaca, UNAM, Mexico}
}

\date{ }

\newtheorem{theo}{Theorem}
\newtheorem{defini}{Definition}
\newtheorem{proposi}{Proposition}
\newtheorem{lemma}{Lemma}
\newtheorem{coro}{Corollary}
\newtheorem{rem}{Remark}

\newcommand{\CM}{{\mathbb C}}

\newcommand{\RM}{{\mathbb R}}
\newcommand{\BM}{{\mathbb B}}
\newcommand{\SM}{{\mathbb S}}

\newcommand{\IM}{{\mathbb I}}
\newcommand{\ZM}{{\mathbb Z}}

\newcommand{\HM}{{\mathbb H}}
\newcommand{\DM}{{\mathbb D}}
\newcommand{\GM}{{\mathbb G}}
\newcommand{\GUM}{{\mathbb G}{\mathbb U}}
\newcommand{\GIM}{{\mathbb G}{\mathbb I}}
\newcommand{\UM}{{\mathbb U}}

\newcommand{\FM}{{\mathbb F}}

\newcommand{\Dd}{{\cal D}}
\newcommand{\Ee}{{\cal E}}
\newcommand{\Ff}{{\cal F}}

\newcommand{\Vv}{{\cal V}}

\newcommand{\Cc}{{\cal C}}

\newcommand{\Kk}{{\cal K}}

\newcommand{\JF}{J}
\newcommand{\JR}{S}

\newcommand{\etaFR}{\tau}
\newcommand{\etaR}{\eta}

\newcommand{\one}{{\bf 1}}

\newcommand{\Uu}{\mathcal{U}}

\newcommand{\diag}{{\mbox{\rm diag}}}
\newcommand{\Ker}{{\mbox{\rm Ker}}}
\newcommand{\Ran}{{\mbox{\rm Ran}}}
\newcommand{\Ind}{{\mbox{\rm Ind}}}
\newcommand{\Sig}{{\mbox{\rm Sig}}}
\newcommand{\Sec}{{\mbox{\rm Sec}}}

\begin{document}

\maketitle

\begin{abstract}
For $J$-hermitian operators on a Krein space $(\Kk,J)$ satisfying an adequate Fredholm property, a global Krein signature is shown to be a homotopy invariant. It is argued that this global signature is a generalization of the Noether index. When the Krein space has a supplementary Real structure, the sets of $J$-hermitian Fredholm operators with Real symmetry can be retracted to certain of the classifying spaces of Atiyah and Singer. Secondary $\ZM_2$-invariants are introduced to label their connected components. Related invariants are also analyzed for $J$-unitary operators. 
\end{abstract}

\vspace{.5cm}

\section{Introduction}
\label{sec-intro}

A Krein space $\Kk$ is a Hilbert space equipped with an indefinite sesquilinear form given by a self-adjoint unitary operator $J$. A bounded linear operator $T$ on $\Kk$ conserving this form in the sense that $T^*JT=J$ is called $J$-unitary. The set of $J$-unitary operators forms a group and its Lie algebra is the set of the $J$-hermitian operators, namely all bounded linear operators $H$ on $\Kk$ satisfying $H^*J=JH$. These two classes of operators can be considered as the equivalent of the unitary and hermitian operators on a Hilbert space, a major difference being though that neither the $J$-unitaries nor the $J$-hermitian operators are necessarily normal. The two classes are further linked via the Cayley transform. Krein spaces as well as $J$-unitary and $J$-hermitian operators thereon have been studied by numerous authors since the 1950's and this is well-documented in the two monographs  \cite{Bog,AI}. Their spectra have reflection properties on the unit circle and the real axis respectively. A further important element of the theory of such operators are the Krein signatures of eigenvalues on the unit circle $\SM^1$ or the real axis $\RM$, given by the signature of $J$ restricted to the associated generalized eigenspaces. Definite signatures imply stability of the eigenvalues, namely they cannot leave the circle or the real line under perturbations. In prior works \cite{Kre,YS,GLR} such signatures were mainly considered for finite dimensional Krein spaces such as the classical groups U$(N,M)$ and SP$(2N,\RM)$, but in \cite{SB} one of the authors thoroughly analyzed Krein signatures for infinite dimensional Krein spaces.

\vspace{.2cm}

This paper considers certain classes of Fredholm operators on Krein spaces. Let us first focus on $J$-hermitian operators. The adequate Fredholm property for them is to require $H-\lambda\,\one$ to be a conventional Fredholm operator for all $\lambda\in\RM$. Such $J$-hermitian operators will be called $\RM$-Fredholm (Definition~\ref{def-essR}). It is shown that the real spectrum of such operators consists of only a finite number of eigenvalues with finite multiplicity (Theorem~\ref{theo-G=F}). Hence these operators can rightfully also be called essentially hyperbolic.  A global homotopy invariant on the set of $J$-hermitian $\RM$-Fredholm operators is then given by the sum of the Krein signatures over all real eigenvalues (Definition~\ref{def-Sigg} and Theorem~\ref{theo-connected}). The global signature invariant replaces the Noether index (often referred to as Fredholm index) in the sense that it labels the connected components of the $J$-hermitian $\RM$-Fredholm operators. In fact, the latter set can be contracted to the Fredholm operators so that the global signature generalizes the Noether index to a wider class of operators (Theorem~\ref{theo-Homotopy}). 

\vspace{.2cm}

In a second step, we consider Krein spaces with Real structures in a spirit similar to Atiyah and Singer \cite{Ati,AS} (Definition~\ref{def-RealKrein}) and then implement these symmetries on the $\JF$-unitary and $\JF$-hermitian operators (Definition~\ref{def-RealUnitaryHermitian}). The large letter R is part of the notation and refers to Atiyah \cite{Ati} who introduced Real $K$-theory to analyze vector bundles that are invariant under a Real symmetry.  In the finite dimensional case, this leads to four classical groups as subgroups of U$(N,M)$ (see Section~\ref{sec-groups}). For the $J$-hermitian $\RM$-Fredholm operators, the Real symmetry leads to restrictions on the values of the global signature and furthermore to the definition of secondary $\ZM_2$-invariants (Theorem~\ref{theo-connectedReal}). The collision scenarios of eigenvalues studied  to show that these invariants are well-defined are also of interest in the finite dimensional cases. In particular, for the classical group O$(N,M)$ a so-called mediated tangent bifurcation is generic and has not been studied elsewhere to our best knowledge. For the infinite dimensional case it is then again shown that the four new classes of $J$-hermitian $\RM$-Fredholm operators with Real symmetry can be retracted to four of the Atiyah-Singer classifying spaces for Real $K$-theory \cite{AS} (Theorem~\ref{theo-connectedReal} and Corollary~\ref{coro-AtiyahSinger}). 

\vspace{.2cm}

For $J$-unitary operators, it is natural to require that $\SM^1$ only contains discrete spectrum, namely to consider the essentially $\SM^1$-gapped $J$-unitaries (Definition~\ref{def-essS1}). This is strictly stronger than imposing  $T-\lambda\,\one$ to be Fredholm for all $\lambda\in\SM^1$. The subtle difference results from the non-normality of the operators involved. On the set of  essentially $\SM^1$-gapped $J$-unitaries the global signature is again a homotopy invariant (Theorem~\ref{theo-SigInv}). This is already discussed in \cite{SB} and is briefly reviewed in Section~\ref{sec-unitaries}. New is, however, the analysis of (secondary) invariants for essentially $\SM^1$-gapped $J$-unitaries with Real symmetries (Theorem~\ref{the-RealInvariants}). It is {\it not} shown that these invariants are complete invariants in the sense that they label the connected components. Indeed, this may not be true. In Section~\ref{sec-unbounded} we conjecture that at least for unbounded $J$-unitaries without Real symmetries (called $J$-isometries as in \cite{Bog}) such a labelling by the global signature is valid. 


\vspace{.2cm}

All the above results are not only of functional analytic interest, but are helpful and relevant for applications of index theory whenever an indefinite sesquilinear form is present. In fact, the augmented flexibility of the new classes of Fredholm operators with global signatures  may then allow to directly define analytic invariants.  An example of such a situation, actually motivating the present work, are the transfer operators for two-dimensional tight-binding models for solid state physics restricted to a half-space. The Krein signatures are then interpreted as the chiralities of the associated boundary states (see Section~6.8 of \cite{SB}). In this framework, the Real symmetries are inherited from physical symmetries of topological insulators. These examples will be studied in detail in a forthcoming work. Preliminary results can be found in \cite{SV}.

\vspace{.2cm}

\noindent {\bf Acknowledgements:} This work considerably extends the mathematical sections of our previous preprint \cite{SV}. Part of the results on $J$-hermitian $\RM$-Fredholm operators were obtained in collaboration with Stefan Daiker during his master thesis in Erlangen in 2014. We thank PAPPIT-UNAM IN 104015 as well as the DFG for financial support.

\section{Invariants for operators on complex Krein spaces}
\label{sec-complexKrein}

Let us recall \cite{Bog,AI,SB} that a Krein space $\Kk$ is a separable complex Hilbert space endowed with a fundamental symmetry, namely a selfadjoint unitary $J$ on $\Kk$. In an adequate basis, the fundamental symmetry takes the form
\begin{equation}
\label{eq-Jchoice}
\JF 
\;=\;
\begin{pmatrix}
\one & \;\;0 \\ 0 & -\one
\end{pmatrix}
\;,
\end{equation}
however, the particular choice of basis will be irrelevant for most arguments in the following. The inertia of $J$ will be denoted by $(N_+,N_-)$, that is $N_\pm$ is the multiplicity of $\pm 1$ as eigenvalue of $J$. We will mainly be interested in the case $N_\pm =\infty$. The term "complex" in the title of this section alludes to the fact that real structures play no role for now, but will later on in Section~\ref{sec-realStruc}. In the following, $\JF $ and its restriction to subspaces of $\Kk$ will often be viewed as hermitian sesquilinear form on $\Kk$ via $(v,w)\mapsto v^*\JF w$. Let us also point out that one can develop the full theory below also for an arbitrary non-degenerate hermitian form, see Remark~\ref{rem-inver} below.

\subsection{Krein signatures for $\JF $-unitaries and $\JF $-hermitians}
\label{sec-KreinSigRev}

\begin{defini} Let $(\Kk,\JF)$ be a Krein space.

\begin{enumerate}[{\rm (i)}]

\item A bounded operator $T\in\BM(\Kk)$ is called $\JF $-unitary if $T^*\JF T=\JF $. The set of $\JF $-unitaries is denoted by
\begin{equation}
\label{eq-Junit0}
\UM(\Kk,\JF )
\;=\;
\{T\in\BM(\Kk)\,:\,T^*\JF  T=\JF \}
\;.
\end{equation}

\item A bounded operator $H\in\BM(\Kk)$ is called $\JF $-hermitian if $\JF H^*\JF =H$.  The set of $\JF $-hermitians is denoted by
\begin{equation}
\label{eq-Jsa}
\HM(\Kk,\JF )
\;=\;
\{H\in\BM(\Kk)\,:\,H^*\JF  =\JF  H\}
\;.
\end{equation}

\end{enumerate}
\end{defini}

If $\Kk$ is finite dimensional, then $\UM(\Kk,\JF )$ is the classical group U$(N_+,N_-)$ and $\HM(\Kk,\JF )$ is isomorphic to its Lie algebra. 

\begin{rem}
\label{rem-inver}
{\rm 
Suppose that $j$ is an invertible selfadjoint on a Hilbert space $\Kk$ and consider the sets $\UM(\Kk,j )$ and $\HM(\Kk,j )$ of $j$-unitary and $j$-hermitian operators defined as in \eqref{eq-Junit0} and \eqref{eq-Jsa}. Then set $J=j|j|^{-1}$. One now has $J^2=\one$ and can check
$$
\UM(\Kk,j ) \;=\;|j|^{-\frac{1}{2}} \UM(\Kk,\JF) |j|^{\frac{1}{2}}\;,
\qquad
\HM(\Kk,j )\;=\;|j|^{-\frac{1}{2}} \HM(\Kk,\JF) |j|^{\frac{1}{2}}
\;.
$$
Therefore it is sufficient to consider the case of $J=J^*$ squaring to the identity. This will be of relevance in Lemma~\ref{lem-transformation} below.
}
\hfill $\diamond$
\end{rem}

Given a bounded linear operator $A\in\BM(\Kk)$, a subset $\Delta\subset\CM$ is called separating if the boundary $\partial\Delta$ has no intersection with the spectrum $\sigma(A)$ of $A$. For a separating set $\Delta\subset\CM$ for $A$, let $P_\Delta(A)$ denote the associated Riesz spectral projection of $A$. Further let us set $\Ee_\Delta(A)=\Ran(P_\Delta(A))$ and $\Ff_\Delta(A)=\Ker(P_\Delta(A))$. If $\Delta=\{\lambda\}$, we also write $P_\lambda(A)$, $\Ee_\lambda(A)$ and $\Ff_\lambda(A)$. If it is clear from the context, the argument $A$ will be dropped. General properties of Riesz projections can be found in \cite{Kat} or the appendix of \cite{SB}. Proofs of the following facts can be found in \cite{Bog,AI,SB}.

\begin{proposi}
\label{prop-basicprops} Let $(\Kk,\JF )$ be a Krein space.

\begin{enumerate}[{\rm (i)}] 

\item $\UM(\Kk,\JF )$ is a group under composition.

\item $\HM(\Kk,\JF )$ is a $\RM$-linear space.

\item For $H\in \HM(\Kk,\JF )$, one has $\exp(\imath \,tH)\in \UM(\Kk,\JF )$ for all $t\in\RM$.

\item For $T\in \UM(\Kk,\JF )$ one has the reflection property $\overline{\sigma(T)}=\sigma(T)^{-1}$ around $\SM^1$. 

\item For $H\in \HM(\Kk,\JF )$ one has the reflection property $\overline{\sigma(H)}=\sigma(H)$ w.r.t. the real axis.

\item For a separating subset $\Delta$ for $T\in \UM(\Kk,\JF )$, one has $P_\Delta(T)^*=\JF P_{(\overline{\Delta})^{-1}}(T)\JF $.

\item For a separating subset $\Delta$ for $H\in \HM(\Kk,\JF )$, one has $P_\Delta(H)^*=\JF P_{\overline{\Delta}}(H)\JF $.

\item For $T\in \UM(\Kk,\JF )$ and $\lambda\in\sigma(T)\cap\SM^1$ a discrete (isolated and of finite multiplicity) eigenvalue,  the hermitian form $\JF  |_{\Ee_\lambda(T)}$  is non-degenerate.

\item For $H\in \HM(\Kk,\JF )$ and $\lambda\in\sigma(H)\cap\RM$ is a discrete eigenvalue,  the hermitian form $\JF  |_{\Ee_\lambda(H)}$  is non-degenerate.

\end{enumerate}
\end{proposi}

By item (iii) the $\JF $-hermitian can be seen as the Lie algebra of $\UM(\Kk,\JF )$. An alternative convention is to require $e^{tB}\in\UM(\Kk,\JF )$ for all $t\in\RM$, which then implies $\JF B^* =-B\JF$ and thus $\sigma(B)=-\overline{\sigma(B)}$. In our opinion this is less practical for the spectral analysis carried out in the present work.
 
\begin{defini} For a discrete  eigenvalue $\lambda\in\sigma(T)\cap\SM^1$ of a $\JF $-unitary $T\in \UM(\Kk,\JF )$ the Krein inertia $\nu(\lambda,T)=(\nu_+(\lambda,T),\nu_-(\lambda,T))$ is the number $\nu_\pm(\lambda,T)$ of positive/negative eigenvalues of the non-degenerate hermitian sesquilinear form $\JF  |_{\Ee_\lambda(T)}$. For eigenvalues off the unit circle the inertia are defined to be $0$. Eigenvalues are called indefinite (sometimes also of mixed signature) if both $\nu_\pm(\lambda,T)$  are non-vanishing, otherwise they are called definite or $\JF$-definite. The Krein signature of $\lambda$ is defined as 
$$
\Sig(\lambda,T)
\;=\;
\nu_+(\lambda,T)\;-\;\nu_-(\lambda,T)
\;.
$$
Similarly, the Krein signature $\nu(\lambda,H)$ and Krein signature $\Sig(\lambda,H)$ of a discrete eigenvalue $\lambda\in\sigma(H)\cap\RM$ of a $\JF $-hermitian $H\in\HM(\Kk,\JF )$ are defined. 
\end{defini}

Let us note that Krein inertia satisfy $\nu_+(\lambda,T)+\nu_-(\lambda,T)=\dim(\Ee_\lambda)$. Moreover, using the Riesz projection, their definition can be rewritten as
\begin{equation}
\label{eq-nuformula}
\nu_\pm(\lambda,T)
\;=\;
\nu_\pm\bigl(P_\lambda^*\,\JF \,P_\lambda\bigr)
\;,
\end{equation}
where on the r.h.s. $\nu_\pm$ denotes the inertia of a sesquilinear form (non-vanishing only on a finite dimensional subspace). The following is the generic bifurcation scenario for eigenvalues to leave the unit circle.

\begin{figure}
\begin{center}
{\includegraphics[width=9cm]{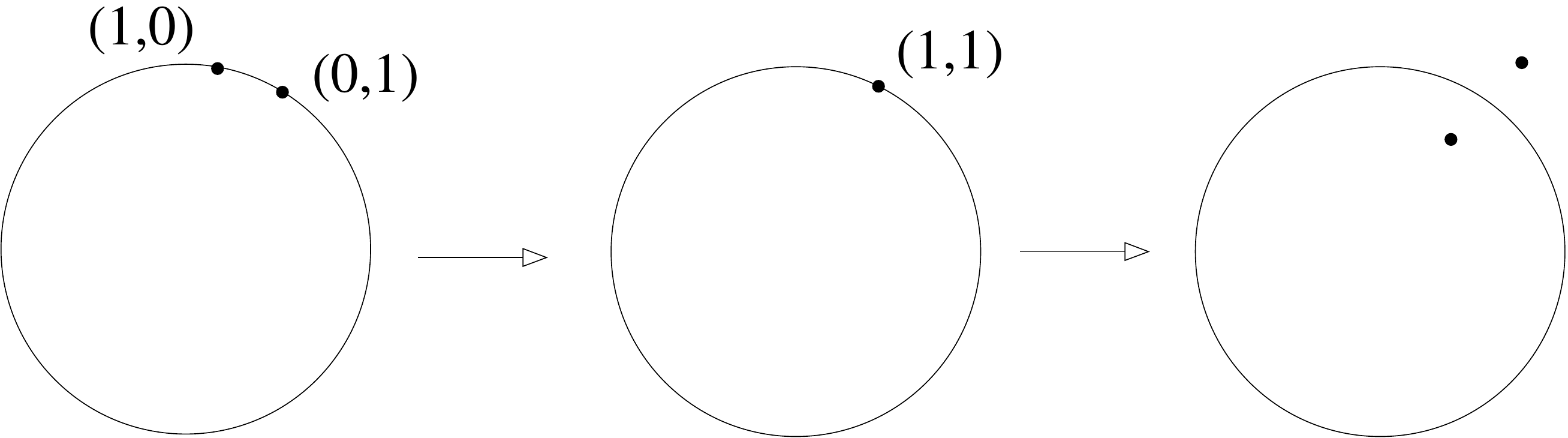}}
\caption{\sl Schematic representation of a Krein collision of eigenvalues with indicated Krein inertia. Note that for eigenvalues with the inertia $(1,0)$ and $(1,0)$ on the l.h.s., one would have inertia $(2,0)$ at the eigenvalue crossing and the Krein stability implies that the eigenvalues cannot leave the unit circle.}
\label{fig-KreinColl}
\end{center}
\end{figure}

\begin{defini}
\label{def-KreinColl}
Given a norm continuous path $t\in[-1,1]\mapsto T_t$ of $\JF $-unitaries, two continuous paths $t\in[-1,1]\mapsto \lambda_{1,2}(t)$ of discrete eigenvalues of $T_t$ undergo a Krein collision at $t=0$ if $\lambda_{1,2}(t)\in\SM^1$ for $t\in(-1,0]$ and $\lambda_{1,2}(t)\not\in\SM^1$ for $t\in(0,1]$, see Figure~\ref{fig-KreinColl}. Similarly, Krein collisions for paths of real discrete eigenvalues of $\JF $-hermitian operators are defined.
\end{defini}
 
Note that even for analytic $t\mapsto T_t$ the colliding eigenvalues do not depend analytically on $t$ (other than for analytic paths of self-adjoint operators). The following stability result goes back to Krein in finite dimension \cite{Kre} and is proved in Section~3.4 of \cite{SB} for $\JF $-unitaries on arbitrary Krein spaces. The proof transposes verbatim to $\JF $-hermitians. 

\begin{theo}[Krein stability]
\label{theo-KreinStab}
Along a norm continuous path $t\mapsto T_t$ of $\JF $-unitaries, discrete eigenvalues can leave the unit circle at $t=0$ by a Krein collision through $\lambda\in\SM^1$ only if both inertia $\nu_+(\lambda,T_0)$ and $\nu_-(\lambda,T_0)$ are non-vanishing. The same statement holds for discrete real eigenvalues of paths of $\JF $-hermitian operators.
\end{theo}

\subsection{Fredholm property and global signature for $\JF $-unitaries}
\label{sec-unitaries}

\begin{defini} 
\label{def-essS1}
A $\JF $-unitary $T\in \UM(\Kk,\JF )$ is said to be  $\SM^1$-Fredholm if $T-\lambda\,\one$ is a Fredholm operator for all $\lambda\in\SM^1$. The set of all $\SM^1$-Fredholm $\JF $-unitaries is denoted by $\FM\UM(\Kk,\JF )$.  A $\JF $-unitary $T\in \UM(\Kk,\JF )$ is said to be essentially $\SM^1$-gapped if there is only discrete spectrum on the unit circle. The set of essentially $\SM^1$-gapped $\JF $-unitaries is denoted by $\GM\UM(\Kk,\JF )$. For $T\in\GM\UM(\Kk,\JF )$, the global signature is defined as
$$
\Sig(T)
\;=\;
\sum_{\lambda\in\SM^1} \Sig(\lambda,T)
\;.
$$
\end{defini}

In \cite{SB} there is an example showing that $\GM\UM(\Kk,\JF )$ is a proper subset of $\FM\UM(\Kk,\JF )$. The same example shows that $\GM\UM(\Kk,\JF )$ is not stable under compact perturbations within the $\JF $-unitaries, a property that clearly holds for $\FM\UM(\Kk,\JF )$ by standard theory of Fredholm operators. Let us also point out that the global signature of $T\in\GM \UM(\Kk,\JF )$ is given by
$$
\Sig(T)\;=\;\Sig\big(\JF |_{\Ee_{\SM^1}(T)}\big)
\;,
$$
where on the r.h.s. appears the well-known signature of sesquilinear form on a finite dimensional vector space. Based on Theorem~\ref{theo-KreinStab} one obtains \cite{SB}:

\begin{theo}
\label{theo-SigInv}
The map $T\in\GM\UM(\Kk,\JF )\mapsto \Sig(T)\in\ZM$ is continuous.
\end{theo}

We do not know whether $\GM\UM_n(\Kk,\JF )=\Sig^{-1}(\{n\})$ is connected or not, but will solve the corresponding question for the $\JF $-hermitian operators in an affirmative way in Theorem~\ref{theo-connected}, and have a further conjecture on this issue for unbounded $\JF$-unitaries in Section~\ref{sec-unbounded}.

\subsection{Fredholm property and global signature for $\JF $-hermitians}
\label{sec-Jhermitian}

\begin{defini} 
\label{def-essR}
A $\JF $-hermitian $H\in \HM(\Kk,\JF )$ is said to be  $\RM$-Fredholm if $H-\lambda\,\one$ is a Fredholm operator for all $\lambda\in\RM$. The set of all $\RM$-Fredholm $\JF $-hermitian operators is denoted by $\FM\HM(\Kk,\JF )$.  
\end{defini}

In view of Definition~\ref{def-essS1}, one may be tempted to introduce the set $\GM\HM(\Kk,\JF )$ of essentially $\RM$-gapped $\JF $-hermitian operators, namely those operators in $\HM(\Kk,\JF )$ which only have discrete spectrum on $\RM$. This is not necessary as one can show $\GM\HM(\Kk,\JF )=\FM\HM(\Kk,\JF )$.
 
\begin{theo}
\label{theo-G=F}
Any $H\in \FM\HM(\Kk,\JF )$ has only discrete spectrum on the real axis.
\end{theo}

Let us stress though that this statement does {\it not} extend to unbounded operators satisfying $H^*J=JH$, namely there are examples (constructed in Section~\ref{sec-unbounded} below) of unbounded $\JF $-self-adjoint $\RM$-Fredholm operators having point spectrum on the whole real axis. The proof of Theorem~\ref{theo-G=F} will be based on the following lemma, which for $a>b$ implies that $\GM\HM(\Kk,\JF )$ is stable under compact perturbations.

\begin{lemma}
\label{lem-DiscStable}
Suppose that $H\in \HM(\Kk,\JF )$ has only discrete spectrum in $\Vv=(-\infty,a)\cup(b,\infty)$. If $K\in  \HM(\Kk,\JF )$ is compact, then also $H+K$ has only discrete spectrum in $\Vv$.
\end{lemma}

\noindent {\bf Proof.} This is an immediate consequence of analytic Fredholm theory ({\it e.g.} Appendix D in \cite{SB}) because $H$ is bounded and both components of $\Vv$ lie in the unbounded component of the resolvent set of $H$.
\hfill $\Box$

\vspace{.2cm}

In several arguments below, we will use the following concept.

\begin{defini}
\label{def-frame}
A frame for a subspace $\Ee\subset\Kk$ is a linear isometric map $\Psi$ from an auxiliary Hilbert space into $\Kk$ such that $\Psi\Psi^*$ is the orthogonal projection on $\Ee$ and $\Psi^*\Psi=\one$.
\end{defini}

\noindent {\bf Proof} of Theorem~\ref{theo-G=F}. Let $H\in \FM\HM(\Kk,\JF )$.  For any $\lambda\in\RM$, one has $\Ker((H-\lambda\,\one)^*)=\JF \,\Ker(H-\lambda\,\one)$ so that $\Ind(H-\lambda\,\one)=0$. Let $\Psi_\lambda$ be a frame for $\Ker(H-\lambda\,\one)$, then $\JF \Psi_\lambda$ is a frame for $\Ker((H-\lambda\,\one)^*)$. Therefore the finite range operator $F_\lambda=\JF \Psi_\lambda\Psi_\lambda^*\in  \HM(\Kk,\JF )$ is such that $H-\lambda\,\one+F_\lambda$ is invertible. Moreover, $H-(\lambda+\epsilon)\,\one+F_\lambda$ is invertible for $|\epsilon|<\epsilon_\lambda$ with $\epsilon_\lambda>0$ sufficiently small. Set $\Uu_\lambda=(\lambda-\epsilon_\lambda,\lambda+\epsilon_\lambda)$. Then $(\Uu_\lambda)_{\lambda\in I}$ is an open cover of the compact set $I=[-\|H\|,\|H\|]$. Let $\lambda_1,\ldots,\lambda_M\in I$ be such that $(\Uu_{\lambda_m})_{m=1,\ldots,M}$ is a finite cover of $I$. Suppose $\lambda_m<\lambda_{m+1}$ and set $F_m=F_{\lambda_m}$ as well as 
$$
\Vv_m\;=\;\Vv_{m-1}\cup \Uu_m\;,
\qquad
\Vv_0\;=\;(-\infty,-\|H\|)\cup(\|H\|,\infty)
\;.
$$
Now $H+F_1$ has no spectrum in $\Uu_1$ by construction (because $H+F_1-\lambda\,\one$ is invertible for all $\lambda\in \Uu_1$), and by Lemma~\ref{lem-DiscStable} actually only discrete spectrum in $\Vv_1$. Applying this lemma to $H+F_1$ and $K=F_2-F_1$, it follows that also $H+F_2$ has only discrete spectrum in $\Vv_1$. Again by construction, $H+F_2$ has no spectrum in $\Uu_2$ and therefore only discrete spectrum in $\Vv_2$. Now iterating this argument $M$ times, it follows that $H+F_M$ has only discrete spectrum in $\Vv_M=\RM$. Invoking again  Lemma~\ref{lem-DiscStable}  it follows that also $H$ has only discrete spectrum on $\RM$.
\hfill $\Box$

\begin{defini}  
\label{def-Sigg}
For $H\in\FM\HM(\Kk,\JF )$, the global Krein signature is defined as
$$
\Sig(H)
\;=\;
\sum_{\lambda\in\RM} \Sig(\lambda,H)
\;.
$$
\end{defini}

By Theorem~\ref{theo-KreinStab} follows 

\begin{theo}
\label{theo-connected}
The map $H\in\FM\HM(\Kk,\JF )\mapsto \Sig(H)\in\ZM$ is continuous.
\end{theo}

Let us give an example of an operator $H\in\FM\HM(\Kk,\JF )$ with signature $n\in\ZM$. For that purpose, let us suppose that $N_+=N_-=\infty$ so that both eigenspaces of $J$ are infinite dimensional. Next choose a Fredholm operator $A:\Ker(\JF -\one)\to\Ker(\JF +\one)$ with Noether index given by $\Ind(A)=n$. Then set, in the grading of $\JF $ given by \eqref{eq-Jchoice},
\begin{equation}
\label{eq-Hspecial}
H
\;=\;
\imath\,
\begin{pmatrix}
0 & A^* \\ A & 0 
\end{pmatrix}
\;.
\end{equation}
One readily checks $H\in\HM(\Kk,\JF )$. Moreover, by construction $H^*=-H$, namely $H$ is skewadjoint. Hence the spectrum of $H$ lies on the imaginary axis. Thus $H\in \FM\HM(\Kk,\JF )$ if and only if $H$ is Fredholm. But this is guaranteed by the Fredholm property of $A$. Furthermore,
$$
\Sig(H)
\;=\;
\Sig(0,H)
\;=\;
\dim(\Ker(A))-\dim(\Ker(A^*))
\;=\;\Ind(A)
\;.
$$
From this perspective, Theorem~\ref{theo-connected} shows that the global signature is an extension  of the Noether index to a larger set of operators. The following result shows that actually the Fredholm operators, identified with the skew-adjoint $\JF $-hermitians via  \eqref{eq-Hspecial}, are a deformation retract of $ \FM\HM(\Kk,\JF )$. 


\begin{theo}
\label{theo-Homotopy}
Suppose that $N_+=N_-$. The set $\FM\HM_n(\Kk,\JF )=\Sig^{-1}(\{n\})$ is connected for all $n\in\ZM$.
\end{theo}

We are mainly interested in the case of infinite and equal $N_\pm$, but for sake of completeness let us state the following complementary result:

\begin{proposi}
\label{prop-Homotopy}
If both $N_\pm$ are finite, then $\FM\HM(\Kk,\JF )=\FM\HM_n(\Kk,\JF )=\Sig^{-1}(\{n\})$ with $n=N_+-N_-$. If one of $N_\pm$ is finite and the other infinite, then $\FM\HM(\Kk,\JF )$ is empty.
\end{proposi}

Theorem~\ref{theo-Homotopy} will be shown by a series of explicit homotopies within $\FM\HM(\Kk,\JF )$. The first one deforms the spectrum of $H$ to three points.

\begin{lemma}
\label{lem-Homotopy1}
$H\in\FM\HM(\Kk,\JF)$ is homotopic to  some $H'\in \FM\HM(\Kk,\JF)$ with $\sigma(H')\subset \{-\imath,0,\imath\}$.
\end{lemma}

\noindent {\bf Proof.}
Let $P_\pm$ be the Riesz projections of $H$ on the spectrum in the upper and lower half-planes $\HM_\pm=\{z\in\CM\,:\,\pm\,\Im m(z)>0\}$, and let $P_0$ be the Riesz projection on the real spectrum of $H$. By Proposition~\ref{prop-basicprops}, $JP_\pm^*J=P_\mp$ and $JP_0^*J=P_0$. Moreover, $\one=P_++P_0+P_-$. Let us set
\begin{equation}
\label{eq-Ht}
H_t\;=\;
(1-t)\,H\;+\;\imath \,t\,(P_+-P_-)
\;,
\qquad
t\in[0,1]
\;.
\end{equation}
By the above relations, one indeed checks $H_t\in \HM(\Kk,J)$. As the ranges of $P_\pm$ and $P_0$ are invariant under $H$, it follows that the spectra of $H_t$ are given by
$$
\sigma(H_t)
\;=\;
\sigma(H_t|_{\mbox{\rm\tiny Ran}(P_+)})\;\cup\;\sigma(H_t|_{\mbox{\rm\tiny Ran}(P_0)})\;\cup\;\sigma(H_t|_{\mbox{\rm\tiny Ran}(P_-)})
\;.
$$
Now by the spectral mapping theorem $\sigma(H_t|_{\mbox{\rm\tiny Ran}(P_\pm)})$ lies in $\HM_\pm$, and $\sigma(H_t|_{\mbox{\rm\tiny Ran}(P_0)})\subset\RM$ is discrete. It follows that $H_t\in\FM\HM(\Kk,J)$. Now $H'=H_1$ is the operator with the desired properties.
\hfill $\Box$
 
\vspace{.2cm}

In order to contract $\FM\HM_n(\Kk,\JF )$ into one point, it will next be necessary to lift the degeneracy of the kernel of $H'$ as far as possible. This will be achieved by an adequate finite dimensional perturbation within $\FM\HM(\Kk,\JF)$. In order to avoid difficulties linked to the perturbation theory of non-normal operators, we will exploit the restriction to invariant subspaces as discussed in the following technical result.

\begin{lemma}
\label{lem-transformation}
Let $\Ee$ be a $\JF$-non-degenerate finite dimensional subspace of $\Kk$ and $\Ff=\JF \Ee^\perp$ be its $\JF$-orthogonal complement where the orthogonal complement $\perp$ is w.r.t. the Hilbert space scalar product on $\Kk$. Let $\Psi$ and $\Phi$ be frames for $\Ee$ and $\Ff$. The idempotents $P_\Ee$ and $P_\Ff$ with ranges $\Ee$ and $\Ff$ as well as kernels $\Ff$ and $\Ee$ respectively, are given by
$$
P_\Ee
\;=\;
\Psi(\Psi^*\JF\Psi)^{-1}\Psi^*\JF
\;,
\qquad
P_\Ff
\;=\;
\Phi(\Phi^*\JF\Phi)^{-1}\Phi^*\JF
\;.
$$
Further set $j_\Psi=\Psi^* \JF \Psi$ and  $n_\Psi=|j_\Psi|^{-\frac{1}{2}}$, and similarly $j_\Phi$ and $n_\Phi$. Then $M=(\Psi \,n_\Psi,\Phi\, n_\Phi)$ is invertible with inverse given by
$$
M^{-1}
\;=\;
\binom{n_\Psi^{-1}\Psi^*P_\Ee}{n_\Phi^{-1} \Phi^*P_\Ff}
\;=\;
\binom{n_\Psi^{-1}j_\Psi^{-1}\Psi^*\JF}{n_\Phi^{-1}j_\Phi^{-1}\Phi^*\JF}
\;.
$$
Further let $H$ be a given linear operator which leaves both $\Ee$ and $\Ff$ invariant. Setting 
\begin{equation}
\label{eq-defJPsi}
H_\Psi\;=\;n_\Psi^{-1}\Psi^*H\Psi\, n_\Psi\;,
\qquad
\JF_\Psi\;=\;n_\Psi^{2}j_\Psi\;=\;j_\Psi\,|j_\Psi|^{-1}
\;,
\end{equation}
and similarly for $H_\Phi$ and $J_\Phi$, one then has $\JF_\Psi^2=\one$ and $\JF_\Phi^2=\one$ and
$$
M^{-1}HM
\;=\;
\begin{pmatrix}
H_\Psi & 0 \\ 0 & H_\Phi
\end{pmatrix}
\;,
\qquad
M^*\JF M
\;=\;
\begin{pmatrix}
\JF_\Psi & 0 \\ 0 & \JF_\Phi
\end{pmatrix}
\;,
$$
and 
$$
H\mbox{ is }\JF\mbox{-hermitian }
\;\;
\Longleftrightarrow
\;\;\;\;
H_\Psi\mbox{ is }\JF_\Psi\mbox{-hermitian }
\;\;\mbox{ and }\;\;
H_\Phi \mbox{ is }\JF_\Phi\mbox{-hermitian }
\;.
$$
Furthermore, the spectra satisfy $\sigma(H)=\sigma(H_\Psi)\cup\sigma(H_\Phi)$.
\end{lemma}

\noindent {\bf Proof.} First of all, $j_\Psi$ is invertible because $\Ee$ is finite-dimensional and not $\JF$-degenerate. Thus $P_\Ee$ is well-defined and it can readily be checked to have the desired properties (see Proposition~2.6 of \cite{SB}). Furthermore, as $\Ee$ is non-degenerate, one has by Lemma~2.3(v) of \cite{SB} that $\Ee\cap\Ff=\{0\}$ and $\Kk=\Ee+\Ff$. As $\Ee$ is finite-dimensional, $\Ee$ and $\Ff$ form a Fredholm pair and thus by Proposition~B.5 of \cite{SB} also $j_\Phi$ is invertible and $P_\Ff$ well-defined with the desired properties. Now by hypothesis, $\Phi^*\JF\Psi=0$ so that $P_\Ee\Phi=0$, $P_\Ff\Psi=0$ and $P_\Ee P_\Ff=P_\Ff P_\Ee=0$. Combining these identities with $\Phi^*\Phi=\one$ and $\Psi^*\Psi=\one$, one readily checks $M^{-1}M=\one $ with the given formulas for $M$ and $M^{-1}$. The $H$-invariance implies $H\Psi=\Psi H_\Psi$ and  $H\Phi=\Phi H_\Phi$ which leads to the formula for $M^{-1}HM$. Furthermore, $H$ is $\JF$-hermitian if and only if $H^*\JF =\JF H$ if and only if $(M^{-1}HM)^*M^*\JF M=M^*\JF MM^{-1}HM$, which when spelled out shows the equivalence. The last claim follows from the formula for $M^{-1}HM$.
\hfill $\Box$
 
\vspace{.2cm}

Lemma~\ref{lem-transformation} can be applied to  $\Ee=\Ran(P_\RM(H))$ and $\Ff=\JF\,\Ker(P_\RM(H)^*)$ whenever $H$ is in $\FM\HM(\Kk,\JF)$. This leads to another proof of Lemma~\ref{lem-Homotopy1}. Our main use will be the following.
 
\begin{lemma}
\label{lem-Homotopy2}
$H\in\FM\HM(\Kk,\JF)$ is homotopic to  some $H''\in \FM\HM(\Kk,\JF)$ with $\sigma(H'')\subset \{-\imath,0,\imath\}$ such that $0$ is a $\JF$-definite eigenvalue of $H''$.
\end{lemma}

\noindent {\bf Proof.} Let us start with $H=H'$ as given in Lemma~\ref{lem-Homotopy1} and use  Lemma~\ref{lem-transformation} for the subspaces $\Ee=\Ee_{0}(H)=\Ran(P_0)$ and $\Ff=J\,\Ker(P_0^*)$ where $P_0=P_\Ee$ is the Riesz projection on the kernel of $H$. Then $H_\Psi=0$. The homotopy will be chosen to be of the form
\begin{equation}
\label{eq-Ht'}
t\in[0,1]\;\mapsto\;
H_t
\;=\;
M
\begin{pmatrix}
t\,V & 0 \\ 0 & H_\Phi
\end{pmatrix}
M^{-1}
\;,
\end{equation}
with a finite dimensional matrix $V$ satisfying $V^*J_\Psi=J_\Psi V$. By Lemma~\ref{lem-transformation} this is a path in $\FM\HM(\Kk,\JF)$ with spectrum given by $\sigma(H_t)=\sigma(H_\Phi)\cup t \,\sigma(V)$. It is also possible to write
\begin{equation}
\label{eq-Ht'bis}
H_t
\;=\;
H\;+\;t\,\Psi\,n_\Psi\, V \,n_\Psi^{-1}\,\Psi^* P_0
\;.
\end{equation}
For the construction of $V$, let the inertia of $J$ on $\Ee$ be $(n_+,n_-)$ and set $n=n_++n_-$. For sake of concreteness, let us suppose $n_+\geq n_->0$ as for $n_-=0$ there is nothing to show. After multiplication from the right by an adequate unitary matrix, the frame $\Psi$ can be chosen such that $\JF_\Psi$ is diagonal with first $n_+$ entries equal to $1$ and then $n_-$ entries equal to $-1$.  Decomposing $\Psi=(\Psi_+,\Psi_0,\Psi_-)$ with $\Psi_\pm$ both spanning $n_-$-dimensional subspaces of $\Ee$, one hence has $\JF_\Psi=\diag(\one_{n_-},\one_{n_0},-\one_{n_-})$ with $n_0=n_+-n_-$. Then set
\begin{equation}
\label{eq-V'choice}
V
\;=\;
\begin{pmatrix}
0 & 0 & \imath \,\one_{n_-}
\\
0 & 0 & 0
\\
\imath\, \one_{n_-} & 0 & 0
\end{pmatrix}
\;,
\end{equation}
It is a matter of calculation to check that $V$ is $J_\Psi$-hermitian. Moreover, the spectrum of $V$ consists of $\imath$ and $-\imath$ both of degenarecy $n_-$, and a kernel of multiplicity $n_0$. By Lemma~\ref{lem-transformation}, the spectrum of $H_t$ is $\{0,\pm \,\imath\,t\}$ and thus $H_1=H''$ has all the desired properties. Let us note that there are many other possibilities to choose $V$.
%
%
%
%
\hfill $\Box$

\vspace{.2cm}

Having a $\JF$-definite kernel as in Lemma~\ref{lem-Homotopy2} now implies that the ranges of the Riesz projections $P_\pm$ are Lagrangian which in turn is of great relevance for the proof of Theorem~\ref{theo-Homotopy} because Lagrangian subspaces are in bijection with unitary operators and can hence be nicely deformed homotopically.

\begin{lemma}
\label{lem-Lag}
Let $H=\imath(P_+-P_-)\in \FM\HM(\Kk,\JF )$ have a $\JF $-definite kernel. Then $\Ee_\pm=\Ran(P_\pm)$ are $\JF $-Lagrangian, namely $\Ee_\pm$ are maximally isotropic for $\JF $.
\end{lemma}

\noindent {\bf Proof.} By Proposition~\ref{prop-basicprops}, $\JF P_\pm^* \JF =P_\mp$ and $\JF P_0^*\JF =P_0$. Therefore $P_\pm^*\JF  P_\pm=\JF P_\mp P_\pm=0$ which implies that $\Ee_\pm$ are isotropic. Now suppose that there is a vector $v\not\in\Ee_+$ such that span$(\Ee_+,v)$ is isotropic. As $P_++P_0+P_-=\one$, we can decompose $v=v_++v_0+v_-$ with $v_\pm\in\Ee_\pm$ and $v_0\in \Ee_0=\Ran(P_0)=\Ker(H)$. Hence either $v_0\not=0$ or $v_-\not = 0$. The isotropy of span$(\Ee_+,v)$ imposes the two conditions $P_+^*\JF v=0$ and $v^*\JF v=0$. From the first condition follows $0=P_+^*\JF v=\JF P_-v=\JF v_-$ so that $v_-=0$ because $\JF $ is invertible. Furthermore, $v_+^*\JF v_0=(P_+v_+)^*\JF (P_0v_0)=v_+^*\JF P_-P_0v_0=0$. Therefore $v^*\JF v=v_0^*\JF v_0$, which has a definite sign for $v_0\not=0$ by the hypothesis that the kernel is definite. This is in contradiction to the second condition $v^*\JF v=0$. Hence $\Ee_+$ is maximally $\JF $-isotropic. The same analysis applies to $\Ee_-$. 
\hfill $\Box$

\vspace{.2cm}

Let us point out that Lemma~\ref{lem-Lag} also holds in finite dimension, but then only the trivial kernel is $\JF $-definite. In infinite dimension, it is possible that both $\Ee_+$ and $\Ee_-$ are Lagrangian, even though they do not span all of $\Kk$. An example for this situation can readily be constructed from \eqref{eq-Hspecial} with $A$ having non-trivial Noether index.

\vspace{.2cm}

\noindent {\bf Proof} of Theorem~\ref{theo-Homotopy}. Let $H=\imath(P_+-P_-)$ be as in Lemma~\ref{lem-Homotopy2}. The aim is to deform $P_\pm$ inside the Fredholm pairs of Lagrangian projections into orthogonal projections. This latter property will assure $H^*=-H$, which when combined with $\JF H^*\JF =H$ leads to the special form \eqref{eq-Hspecial}. This completes the proof because it is known that the connected components of the set of Fredholm operators are labelled by the Noether index. First of all, let us express the Riesz projections $P_\pm$ in terms of frames $\Phi_\pm$ for $\Ran(P_\pm)$:
\begin{equation}
\label{eq-PRep}
P_\pm
\;=\;
\Phi_\pm(\Phi_\mp^*\JF \Phi_\pm)^{-1}\Phi_\mp^*\JF 
\;.
\end{equation}
One readily checks that the r.h.s. is idempotent and has the correct range and kernel. The existence of the inverse is proved along the lines of Propositions 5.12 and 5.13 of \cite{SB}, which readily transposes from $\JF $-unitaries to $\JF $-hermitians. For the convenience of the reader, let us sketch the argument. The spaces $\Ee_\pm=\Ran(P_\pm)$ are closed and have trivial intersection, and furthermore $\Ee_+\oplus\Ee_-$ has finite co-dimension. Thus $\Ee_+$ and $\Ee_-$ form a Fredholm pair. Because $\Ker(P_\pm)$ differs from $\Ee_\mp$ only by a finite dimensional subspace, also $\Ee_\pm$ and $\Ker(P_\mp)=\JF \,\Ran(P_\pm)^\perp$ form a Fredholm pair. But $\JF \Phi_\pm$ is a frame for $\JF \,\Ran(P_\pm)$ and therefore general Hilbert space principles (Proposition~B.5 of \cite{SB}) lead to the above formula. The Fredholm property is actually equivalent to the invertibility of $\Phi_\mp^*\JF \Phi_\pm$ by Theorem~B.4 of \cite{SB}.

\vspace{.1cm}

Now we use the fact $\Ee_\pm$ are Lagrangian by Lemma~\ref{lem-Lag}. As $N_+=N_-$, Theorem~2.10 of \cite{SB} implies that it is possible to choose the frames to be of the form
\begin{equation}
\label{eq-LagU}
\Phi_\pm
\;=\;
\frac{1}{\sqrt{2}}\,
\begin{pmatrix}
u_\pm \\ \one
\end{pmatrix}
\;,
\end{equation}
with two unitaries $u_\pm$ and where the block entries are given in the grading of $\JF $ given by \eqref{eq-Jchoice}. Now $P_\pm^*=P_\pm$ is guaranteed if $\Phi_\pm=-\JF \Phi_\mp$, namely $u_\pm=-u_\mp$. This will be achieved by a homtopy $t\in[0,1]\mapsto u_{\pm,t}$ from $u_{\pm,0}=u_\pm$ to $u_{+,1}=u_{-,1}$ along which the invertibility of $u_-^*u_+-\one$ is insured, so that the associated Lagrangian frames $\Phi_{\pm,t}$ defined as in \eqref{eq-LagU} always have non-intersecting ranges forming a Fredholm pair. For that purpose, let us choose a path $t\in[0,1]\mapsto v_t$ of unitaries with $1\not\in\sigma(v_t)$ such that $v_0=u_-^*u_+$ and $v_1=-\one$. The existence of such a path is guaranteed by spectral calculus. Then set $u_{+,t}=u_+$ and $u_{-,t}=u_+v_t^*$. This path satisfies all the desired properties.
\hfill $\Box$

\vspace{.2cm}

\noindent {\bf Proof} of Proposition~\ref{prop-Homotopy}. Lemmas~\ref{lem-Homotopy1}, \ref{lem-Homotopy2} and \ref{lem-Lag} remain valid and are applied first. The dimension of any Lagrangian plane is $\min\{N_-,N_+\}$, and therefore also of $\Ee_\pm$. For finite $N_\pm$, this implies that the $\JF$-definite kernel is of dimension $N_++N_--\min\{N_-,N_+\}$. This fixes the signature. If exactly one of $N_\pm$ is infinite, then $P_0$ has to be infinite dimensional and the operator is not $\RM$-Fredholm.
\hfill $\Box$

\subsection{Cayley transforms mapping $\JF $-unitaries to $\JF $-hermitians}
\label{sec-Cayley}

In this section, the global signatures of $\JF $-unitaries and $\JF $-hermitians are connected via the Cayley transform. Furthermore the Cayley transform is used in Sections~\ref{sec-InvJherm} and \ref{sec-unbounded}. For $z\in\CM\setminus\RM$ and $\zeta\in\SM^1$, the Cayley transform $C_{z,\zeta}$ and its inverse are defined by
$$
C_{z,\zeta}(\lambda)
\;=\;
\zeta\,
\frac{\lambda-z}{\lambda-\overline{z}}\;,
\qquad
C^{-1}_{z,\zeta}(\lambda)
\;=\;
\frac{z\,\zeta-\overline{z}\,\lambda}{\zeta-\lambda}
\;.
$$
These are viewed as maps on the Riemann sphere $\overline{\CM}=\CM\cup\{\infty\}$. Note that for real $z$ these formulas still make sense, but define uninteresting maps.  Let us list a few elementary mapping properties:

\begin{enumerate}[(i)]

\item $C_{z,\zeta}:\overline{\RM}\to\SM^1$ is a bijection (with $\overline{\RM}$ being the one-point compactification of $\RM$)

\item $C_{z,\zeta}(z)=0$, $C_{z,\zeta}(\overline{z})=\infty$, $C_{z,\zeta}(\infty)=\zeta$ and  $C_{z,\zeta}(0)=\zeta\,\frac{z}{\overline{z}}$

\item $C_{z,\zeta}\circ C_{z,\zeta}^{-1}=\,$id

\item For $z\in\HM_\pm=\{z\in\CM\,:\,\pm\Im m(z)>0\}$, the map $C_{z,\zeta}:\HM_\pm\to\DM=\{z\in\CM\,:\,|z|<1\}$ is a bijection.

\end{enumerate}

\begin{proposi}
\label{prop-Cayley} For $z\in\CM\setminus\RM$ and $\zeta\in\SM^1$, set 
$$
\HM_z(\Kk,\JF )
\;=\;
\{H\in \HM(\Kk,\JF )\;|\;z\not\in\sigma(H)\}\;,
\qquad
\UM_\zeta(\Kk,\JF )
\;=\;
\{T\in \UM(\Kk,\JF )\;|\;\zeta\not\in\sigma(T)\}\;.
$$
Then one has injections
$$
C_{z,\zeta}:\HM_z(\Kk,\JF )\to\UM(\Kk,\JF )
\;,
\qquad
C_{z,\zeta}(H)\;=\;\zeta\,(H-z\,\one)(H-\overline{z}\,\one)^{-1}\;,
$$
and
$$
C_{z,\zeta}^{-1}:\UM_\zeta(\Kk,\JF )\to\HM(\Kk,\JF )
\;,
\qquad
C^{-1}_{z,\zeta}(T)\;=\;(z\zeta\,\one-\overline{z}\,T)(\zeta\,\one-T)^{-1}\;.
$$
Furthermore, for $\FM\HM_z(\Kk,\JF )=\FM\HM(\Kk,\JF )\cap\HM_z(\Kk,\JF )$ and $\GM\UM_\zeta(\Kk,\JF )=\GM\UM(\Kk,\JF )\cap\UM_\zeta(\Kk,\JF )$ one has
$$
C_{z,\zeta}:\FM\HM_z(\Kk,\JF )\to\GM\UM(\Kk,\JF )
\;,
\qquad
C_{z,\zeta}^{-1}:\GM\UM_\zeta(\Kk,\JF )\to\FM\HM(\Kk,\JF )
\;.
$$
\end{proposi}

The proofs of all statements are straightforward, but can also be found in Section VI.8 of \cite{Bog}. Moreover, it is shown in \cite{Bog} that the spectral mapping properties under Cayley transform hold for all components of the spectrum (point, discrete, continuous, residual).  Because $\Ee_{C_{z,\zeta}(\lambda)}(C_{z,\zeta}(H))=\Ee_\lambda(H)$ for $H\in\FM\HM_z(\Kk,\JF )$ whenever everything is well-defined, the Cayley transform also maps Krein inertia of eigenvalues via
\begin{equation}
\label{eq-InertiaCaley}
\nu_\pm(\lambda,H)
\;=\;
\nu_\pm(C_{z,\zeta}(\lambda),C_{z,\zeta}(H))
\;,
\qquad
\nu_\pm(\lambda,T)
\;=\;
\nu_\pm(C^{-1}_{z,\zeta}(\lambda),C^{-1}_{z,\zeta}(T))
\;.
\end{equation}
This implies the following global statement.

\begin{proposi}
\label{prop-CayleySig} For $H\in\FM\HM_z(\Kk,\JF )$ and $T\in \GM\UM_\zeta(\Kk,\JF )$, one has
$$
\Sig(H)\;=\;\Sig(C_{z,\zeta}(H))
\;,
\qquad
\Sig(T)\;=\;\Sig(C^{-1}_{z,\zeta}(T))
\;.
$$
\end{proposi}

\section{Invariants for operators on Real Krein spaces}
\label{sec-realStruc}

Let us recall that a real structure on the complex vector space  $\Kk$ is an anti-linear involution, namely a map $\Cc:\Kk\to\Kk$ with $\Cc(v+\lambda\, w)=\Cc(w)+\overline{\lambda}\,\Cc(v)$ for all $\lambda\in\CM$ and $\Cc^2=\one$. We will simply use the notation $\overline{v}=\Cc v$. Furthermore, given a linear operator $A$ on $\Kk$, the operator $\Cc A\Cc$ is also a linear operator which we denote by $\overline{A}$. An operator is then called real if $\overline{A}=A$.

\begin{defini} 
\label{def-RealKrein}
A Real Krein space of kind $(\etaR,\etaFR)\in\{-1,1\}^2$ is a complex Krein space $(\Kk,\JF)$ with a Real fundamental symmetry $\JF=\overline{\JF}$ together with a second real symmetry operator $\JR=\overline{\JR}$ satisfying 
$$
\JR^2\;=\;\etaR\,\one\;,
\qquad
\JF\,\JR\,=\,\etaFR\,\JR\,\JF
\;.
$$
\end{defini}

For $(\eta,\tau)=(1,-1)$, this means that $J$ and $S$ provides a representation on $\Kk$ of the real Clifford algebra $Cl_{2,0}$, and for $(\eta,\tau)=(-1,-1)$ of $Cl_{1,1}$. It will  be necessary to use an adequate representation bringing both $\JF$ and $\JR$ into their normal form. This can be achieved by an orthogonal basis transformation, {\it e.g.} Proposition~13 of \cite{GS}. The outcome is that $\JF$ is of the form \eqref{eq-Jchoice} and for the kinds $(\etaR,1)$ and $(\eta,-1)$  respectively, $\JR$ is given by
\begin{equation}
\label{eq-Schoice}
\JR
\;=\;
\begin{pmatrix}
\JR_+ & \;\;0 \\ 0 & \JR_-
\end{pmatrix}
\;\;\;
\mbox{\rm with }\;(\JR_\pm)^2=\etaR\,\one
\;,
\qquad
\JR
\;=\;
\begin{pmatrix}
0 & \eta\,\one \\ \one & 0
\end{pmatrix}
\;.
\end{equation}
Let us point out that in the two cases $\tau=-1$, this implies that the two eigenspaces of $\JF$ are of same dimension. For $\tau=1$, there is more freedom. We only restrict to particular cases in Theorem~\ref{theo-connectedReal} below.

\subsection{Groups and algebras of operators on Real Krein spaces}
\label{sec-groups}

\begin{defini} 
\label{def-RealUnitaryHermitian}
Let $(\Kk,\JF,\JR)$ be a Real Krein space of kind $(\etaR,\etaFR)\in\{-1,1\}^2$.

\begin{enumerate}[\rm (i)]

\item The set of $\JF$-unitaries with Real symmetry $\JR$ is defined by
\begin{equation}
\label{eq-RSym}
\UM(\Kk,\JF,\JR)
\;=\;
\left\{
T\in\UM(\Kk,\JF)\;\left|\;\JR^*\,\overline{T}\,\JR=T\right.\right\}
\;.
\end{equation}

\item The set of $\JF$-hermitians with Real symmetry $\JR$ is defined by
\begin{equation}
\label{eq-RSymHermitian}
\HM(\Kk,\JF,\JR)
\;=\;
\left\{
H\in\HM(\Kk,\JF)\;\left|\;\JR^*\,\overline{H}\,\JR=-H\right.\right\}
\;.
\end{equation}

\end{enumerate}

\end{defini}

The set $\UM(\Kk,\JF,\JR)$ forms a subgroup of  $\UM(\Kk,\JF)$, and $\HM(\Kk,\JF,\JR)$ is isomorphic to its Lie algebra. Note that $\exp(\imath\, t H)\in\UM(\Kk,\JF,\JR)$ for $H\in\HM(\Kk,\JF,\JR)$ as in Propostion~\ref{prop-basicprops}(iii) only due to the minus sign in \eqref{eq-RSymHermitian}. 

\vspace{.2cm}

\noindent {\bf Examples} Let us write out explicitly the finite dimensional examples of the groups in Definition~\ref{def-RealUnitaryHermitian} corresponding to the four possible choices of the signs $\etaR$ and $\etaFR$. Hence let $\Kk=\CM^{N_++{N_-}}$ and $J=\diag(\one_{N_+},-\one_{N_-})$. There are four different subgroups of U$(N_+,{N_-})$ listed in Table~1. For $\etaR=\etaFR=1$, one can choose $\JR=\one$. Then the reality relation in \eqref{eq-RSym} immediately leads to $ \mbox{\rm O}(N_+,{N_-})$. For $\etaR=-1$ and $\etaFR=1$, one needs $S$ commuting with $J$ and squaring to minus the identity. This forces $J$ to have even-dimensional eigenspaces, so we choose $J=\diag(\one_{N_+},-\one_{N_-})\otimes\one_2$ and $S=\one_{{N_+}+{N_-}}\otimes \binom{0 \;-1}{1\;\;\;0}$. Then the reality relation in \eqref{eq-RSym} leads directly to the definition of $\mbox{\rm SP}(2{N_+},2{N_-})$. For the remaining two cases $(\etaR,-1)$, the anti-commutation relation implies that ${N_-}={N_+}$. Then $\JF=\diag(\one,-\one)$ and  $\JR=\binom{0 \;\eta\one}{\one\;\;0}$ with all block entries of size $N_+$. Writing out the corresponding relations in $\UM(\Kk,\JF,\JR)$ does not lead directly to the defining  relations of  $\mbox{\rm SP}(2{N_+},\RM)$ and $\mbox{\rm SO}^*(2{N_+})$ respectively, but after a Cayley transform precisely these relations become apparent. See also \cite{GS} for details.
\hfill $\diamond$

\begin{table*}
\label{table1}
\begin{center}
\begin{tabular}{|c|c||c||c||c||c||c|}
\hline
$\etaR$ &$\etaFR$  & Class. $\!\!\!\!$ Group  & Inertia & Bifurcation & $\pi_0\supset$ & Invariant  
\\
\hline\hline
&  & $\mbox{\rm U}({N_+},{N_-})$ & & KC & $\ZM$ & $\Sig$ 
\\
\hline\hline
$1$ & $1$ &  $\mbox{\rm O}({N_+},{N_-})$ & $\nu_\pm(\lambda)=\nu_\pm(\overline{\lambda})$ & $\!$QKC, MTB, MPD$\!$ & $\ZM\times\ZM_{2}$ & $\Sig\times\Sec$ 
\\
\hline
$-1$ & $ -1$  & $\mbox{\rm SO}^*(2{N_+})$ & $\nu_\pm(\lambda)=\nu_\mp(\overline{\lambda})$ & QKC & $\ZM_2$ & $\Sig_2$ 
\\
\hline
$-1$ & $1$  & $\!\mbox{\rm SP}(2{N_+},2{N_-})\!$  & $\nu_\pm(\lambda)=\nu_\pm(\overline{\lambda})$ & QKC & $2\,\ZM$ & $\Sig$
\\
\hline
$1$ & $-1$ & $\mbox{\rm SP}(2{N_+},\RM)$ & $\nu_\pm(\lambda)=\nu_\mp(\overline{\lambda})$ & QKC, TB, PD & $1$ & 
\\
\hline
\end{tabular}
\caption{\sl In dependence on the kind $(\etaR,\etaFR)$ of the Real Krein space are listed: the classical groups obtained from $\UM(\Kk,\JF,\JR)$ in the case of a finite dimensional Krein space; the symmetry of the inertia of eigenvalues on the unit circle; the bifurcations of eigenvalues on the unit circle in the terminology of Definition~\ref{def-RealKreinColl}; the minimal number of connected components $\pi_0=\pi_0(\GUM(\Kk,\JF,\JR))$; the global invariant labeling these components constructed in Section~\ref{sec-KreinRealSig}.
}
\end{center}
\end{table*}

\vspace{.2cm}

\subsection{Spectral properties of operators with Real symmetry}
\label{sec-spectral}

The following proposition collects the basic spectral implications of the Real symmetry. 

\begin{proposi}
\label{prop-JunitaryTRI} Let $T\in \UM(\Kk,\JF,\JR)$ and $H\in\HM(\Kk,\JF,\JR)$. 

\begin{enumerate}[\rm (i)]

\item The spectra satisfy
$$
\sigma(T)
\;=\;
\overline{\sigma(T)}^{-1}
\;=\;
\overline{\sigma(T)}
\;=\;
\sigma(T)^{-1}
\;,
\quad
\sigma(H)
\;=\;
\overline{\sigma(H)}
\;=\;
-\,\overline{\sigma(H)}
\;=\;
-\,\sigma(H)
\;.
$$
In particular, discrete eigenvalues of $T$ always come either in quadruples or in couples lying on $\SM^1$, and those of $H$ either quadruples or couples on $\RM$.

\item $S^*\,\overline{P_\Delta(T)}\,S\,=\,P_{\overline{\Delta}}(T)$ and $S^*\,\overline{P_\Delta(H)}\,S\,=\,P_{-\overline{\Delta}}(H)$ for any separating set $\Delta\subset \CM$.

\item  If $T \psi=\lambda \psi$, then $T (\JR \,\overline{\psi})=\overline{\lambda}(\JR\,\overline{\psi})$. If $H\psi=\lambda \psi$, then $H(\JR \,\overline{\psi})=-\,\overline{\lambda}(\JR\,\overline{\psi})$. 

\item {\rm (Kramers degeneracy)} Let $\etaR=-1$.  Real eigenvalues of $T\in \UM(\Kk,\JF,\JR)$ and purely imaginary eigenvalues of $H\in \HM(\Kk,\JF,\JR)$  have even geometric and algebraic multiplicity.

\item For all discrete eigenvalues $\lambda\in\sigma(T)\cap\SM^1$, one has $\nu_\pm(\lambda,T)=\nu_{\pm\etaFR}(\overline{\lambda},T)$. Similarly, for all discrete eigenvalues  $\lambda\in\sigma(H)\cap\RM$, one has $\nu_\pm(\lambda,H)=\nu_{\pm\etaFR}(-\lambda,H)$. 

\item For $z\in\imath\RM$ and $\zeta\in\{-1,1\}$, $C_{z,\zeta}(H)\in \UM(\Kk,\JF,\JR)$ and $C^{-1}_{z,\zeta}(T)\in\HM(\Kk,\JF,\JR)$.

\end{enumerate}
\end{proposi}

\noindent {\bf Proof.} We only consider the case of the $J$-unitary operator, as the case of a $J$-hermitian is similar. (i) The first equality already holds for $\JF$-unitaries without Real symmetry. The second follows from the identity $\JR^*\overline{(T-\overline{z}\,\one})\JR=T-z\,\one$ and the invertibility of $\JR$. Using the first identity, this then also implies the last identity. For (ii), one rewrites the Riesz projections $P_\Delta=P_\Delta(T)$ using the positively oriented path $\Gamma_\Delta$ around $\Delta$:
$$
\JR^*\,\overline{P_\Delta}\,\JR
\;=\;
\oint_{\overline{\Gamma_\Delta}} \frac{dz}{2\pi(-\imath)}\;
\JR^*\,(z\,\one- \overline{T})^{-1}\,\JR
\;=\;
\oint_{\Gamma_{\overline{\Delta}}} \frac{dz}{2\pi\imath}\;
(z\,\one- \JR^*\overline{T}\JR)^{-1}
\;=\;
P_{\overline{\Delta}}
\;.
$$
Here $\overline{\Gamma_\Delta}$ is a negatively oriented path around $\overline{\Delta}$, and the positively oriented path with same graph is denoted by $\Gamma_{\overline{\Delta}}$. (iii) is readily checked. (iv) The algebraic multiplicity of a discrete eigenvalue $\lambda$ is the dimension of the range of its Riesz projection. This range satisfies $\Ee_\lambda=\JR\,\overline{\Ee_\lambda}$. Hence Lemma~\ref{lem-Kramers} below can be applied to conclude. In a similar way, the geometric multiplicity is treated invoking (iii). Using $\overline{\JF}=\JF$, Sylvester's theorem and $\JR\JF\JR^*=\etaFR\JF$, one finds from \eqref{eq-nuformula}
\begin{eqnarray*}
\nu_\pm(\lambda,T)
& = &
\nu_\pm(P_\lambda^*\,\JF P_\lambda)
\;=\;
\nu_\pm\bigl(\,\overline{P_\lambda^*\,\JF P_\lambda}\bigr)
\;=\;
\nu_{\pm}\bigl(\,\overline{P_\lambda^*}\,\JF \overline{P_\lambda}\bigr)
\\
& = &
\nu_{\pm}\bigl(\JR^*P_{\overline{\lambda}}^*\JR\,\JF\JR^*P_{\overline{\lambda}}\JR\bigr)
\;=\;
\nu_{\pm\etaFR}\bigl(P_{\overline{\lambda}}^*\,\JF P_{\overline{\lambda}}\bigr)
\;=\;
\nu_{\pm\etaFR}(\overline{\lambda},T)
\;.
\end{eqnarray*}
This shows (v). The last item is a short calculation.
\hfill $\Box$

\vspace{.2cm}

\begin{figure}
\begin{center}
{\includegraphics[width=9cm]{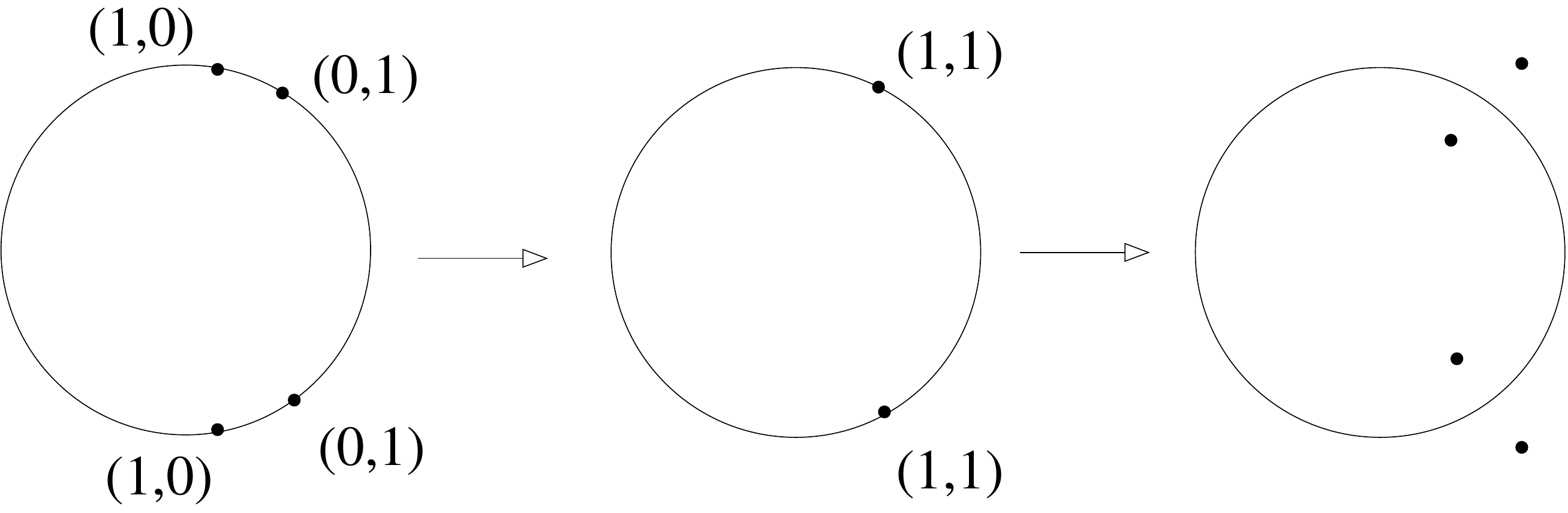}}
\caption{\sl Schematic representation of a quadruple Krein collision of eigenvalues with indicated Krein inertia in the case $\etaFR=1$ where $\nu_\pm(\lambda,T)=\nu_\pm(\overline{\lambda},T)$. For $\etaFR=-1$ the quadruple Krein collision looks quite similar, only the Krein inertia of the pair on either the upper or the lower part of the circle have to be adequately modified.}
\label{fig-KreinCollQuad}
\end{center}
\end{figure}

\begin{lemma}
\label{lem-Kramers}
Let $\etaR=-1$. If $\Ee\subset\Kk$ is a finite dimensional subspace satisfying $\JR\,\overline{\Ee}=\Ee$, then $\dim(\Ee)$ is even. 
\end{lemma}

\noindent {\bf Proof.} 
We construct iteratively subpaces $\Ee_{2n}=\mbox{\rm span}\{\psi_1,\JR\,\overline{\psi_1},\ldots,\psi_n,\JR\,\overline{\psi_n}\}$ of $\Ee$ of dimension $2n$. For $n=1$ choose some $\psi_1\in\Ee$. Suppose that $\psi_1=\mu \JR\, \overline{\psi_1}$ for some $\mu\not=0$. Then $\psi_1=\mu \JR\,\overline{(\mu \JR\, \overline{\psi_1})}=-|\mu|^2 \psi_1$ implying $\psi_1=0$. Hence either $\Ee$ is trivial or at least two-dimensional. Now let $\Ee_{2n}$ as above be given. Choose $\psi_{n+1}\in\Ee\setminus\Ee_{2n}$. By hypothesis, $\JR\,\overline{\psi_{2n+1}}\in\Ee$. Now define $\Ee_{2n+2}$ by the above formula. If $\dim(\Ee_{2n+2})=2n+2$, one continues the iterative construction. If $\dim(\Ee_{2n+2})=2n+1$, then $\JR\,\overline{\psi_{n+1}}=\mu\psi_{n+1}+\phi$ for some $\phi\in\Ee_{2n}$ and $\mu\in\CM$. Hence with $\JR\,{\psi_{n+1}}=\overline{\mu}\,\overline{\psi_{n+1}}+\overline{\phi}$ and $\JR^2=-\one$ implies
$$
\JR\,\overline{\psi_{n+1}}
\;=\;
-\mu\JR(\overline{\mu}\,\overline{\psi_{n+1}}+\overline{\phi})
\,+\,\phi
\;=\;
-|\mu|^2\JR\,\overline{\psi_{n+1}}\,-\,\mu\JR\,\overline{\phi}\,+\,\phi
\;.
$$
Now $\phi-\mu\JR\,\overline{\phi}\in\Ee_{2n}$ so that also $\JR\,\overline{\psi_{n+1}}\in\Ee_{2n}$. Using again that $\JR\,\overline{\Ee_{2n}}=\Ee_{2n}$ follows that also $\psi_{n+1}\in\Ee_{2n}$, in contradiction to the above choice. In conclusion, either $\Ee=\Ee_{2n}$ or $\Ee_{2n+2}\subset\Ee$. Hence the dimension of $\Ee$ is even. 
\hfill $\Box$

\vspace{.2cm}

\begin{figure}
\begin{center}
{\includegraphics[width=9cm]{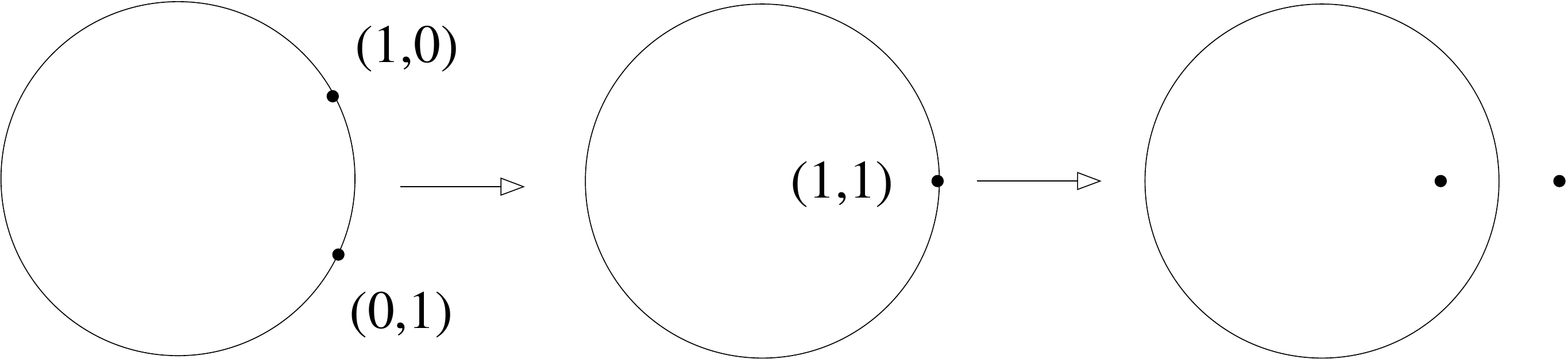}}
\caption{\sl Schematic representation of a tangent bifurcation of eigenvalues with indicated Krein inertia.}
\label{fig-TangBir}
\end{center}
\end{figure}

The Krein stability analysis of eigenvalues on the unit circle for $J$-unitaries (and on the real axis for $J$-hermitians) still holds in presence of symmetries. However, the spectral symmetries stated in Proposition~\ref{prop-JunitaryTRI} also affect the Krein collisions. 
In particular, the reflection symmetric points $1$ and $-1$ of the unit circle play a special role for $J$-unitaries, and $0$ a special role for $J$-hermitians. For real symplectic matrices, the generic destabilization routes for eigenvalues on the unit circle are well-known \cite{HM}. This corresponds to $J$-unitaries on a Real Krein space of kind $(1,-1)$, but in one of the other cases new bifurcation scenarios are generic. 

\begin{figure}
\begin{center}
{\includegraphics[width=9cm]{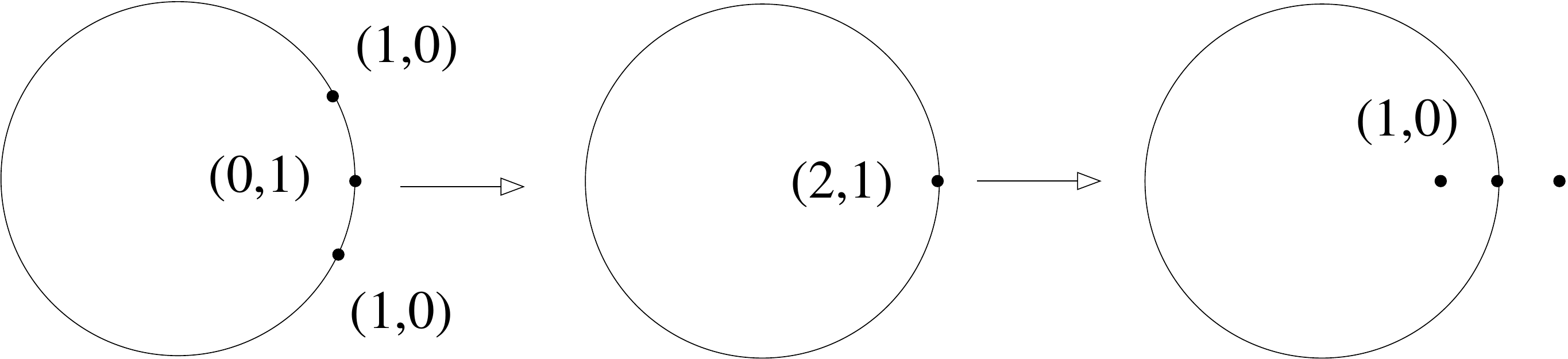}}
\caption{\sl Schematic representation of a mediated tangent bifurcation of eigenvalues with indicated Krein inertia.}
\label{fig-TangBirMed}
\end{center}
\end{figure}

\begin{defini}
\label{def-RealKreinColl}
Let $(\Kk,\JF,\JR)$ be a Real Krein space and $t\in[-1,1]\mapsto T_t\in\UM(\Kk,\JF,\JR)$ a norm continuous path, and let there be continuous paths $t\in[-1,1]\mapsto \lambda_{1,2}(t)$ of discrete eigenvalues of $T_t$ undergoing a Krein collision (KC) at $t=0$ through $\lambda_0=\lambda_{1,2}(0)\in\SM^1$. The KC is called

\begin{enumerate}[\rm (i)]

\item  a quadruple Krein collision (QKC)  if $\lambda_0\not\in\{-1,1\}$, see Figure~\ref{fig-KreinCollQuad}; 

\item a tangent bifurcation (TB) if $\lambda_0=1$ is of multiplicity $2$, see Figure~\ref{fig-TangBir}; 

\item a mediated tangent bifurcation (MTB) if $\lambda_0=1$ is of multiplicity $3$, see Figure~\ref{fig-TangBirMed}; 

\item a period doubling bifurcation (PD) if $\lambda_0=-1$ is of multiplicity $2$;

\item a mediated period doubling bifurcation (MPD) if $\lambda_0=-1$ is of multiplicity $3$.

\end{enumerate}
%
\end{defini}

\begin{proposi}
\label{prop-scenarios} Let $(\Kk,\JF,\JR)$ be a Real Krein space of kind $(\etaR,\etaFR)$.

\begin{enumerate}[\rm (i)]

\item For $(\etaR,\etaFR)=(1,1)$, the generic destabilization scenarios are QKC, MTB, MPD.

\item For $(\etaR,\etaFR)=(-1,-1)$, the generic destabilization scenario is QKC.

\item For $(\etaR,\etaFR)=(-1,1)$, the generic destabilization scenario is QKC.

\item For $(\etaR,\etaFR)=(1,-1)$, the generic destabilization scenarios are QKC, TB, PD.

\end{enumerate}
\end{proposi}

\noindent {\bf Proof.}  The main tools in the proof will be Kramers degeneracy and the reflection symmetry $\nu_\pm(\lambda,T)=\nu_{\pm\etaFR}(\overline{\lambda},T)$ of the intertia as proved in Proposition~\ref{prop-JunitaryTRI}(iv) and (v). This does not affect the QKC which is hence generic in all cases. However, it is relevant for the other scenarios. (i) As $\etaFR=1$, the two simple eigenvalues of a TB or PD have equal inertia before the collision, and therefore lead to a definite eigenvalue at the collision. By Theorem~\ref{theo-KreinStab}, no such eigenvalue can leave the unit circle. Thus there is no TB and PD possible. On the other hand, if $\pm 1$ are simple eigenvalues having an opposite inertia w.r.t. the colliding eigenvalues, the eigenvalue at the collision becomes indefinite. Then it is possible for two of the eigenvalues to leave the unit circle. This are the MTB and MPB. See Figure~\ref{fig-TangBirMed} for an illustration. (iii)  As still $\etaFR=1$,  the TB and PD are suppressed by the above argument. Because now $\etaR=-1$, the eigenvalues $\pm 1$ have Kramers degeneracy. If the inertia of this eigenvalue is $(1,1)$, it generically leaves the circle before the collision with the other two eigenvalues. If the eigenvalue is definite, a generic perturbation will drive them out of $\pm 1$ and afterwards one would again have a QKC. (iv)  Now $\nu_\pm(\lambda,T)=\nu_{\mp}(\overline{\lambda},T)$ and the TB and PD are possible and generic. As in particular,  $\nu_+(\pm 1,T)=\nu_{-}(\pm 1,T)$, the MTB and MPD are not possible. (ii) By the same argument, MTB and MPD are suppressed. Furthermore, TB and PD are not possible because all real eigenvalues have Kramers degeneracy, so the two eigenvalues $\pm 1$ of multiplicity $2$ cannot split into two real eigenvalues, and also not into a quadruple of the unit circle.
\hfill $\Box$

\subsection{Invariants for gapped $J$-unitaries with Real symmetry}
\label{sec-KreinRealSig}

The main focus of this section will be on the sets 
$$
\GUM(\Kk,\JF,\JR)
\;=\;
\GUM(\Kk,\JF)\,\cap\,\UM(\Kk,\JF,\JR)
\;,
$$ 
of essentially $\SM^1$-gapped $\JF$-unitaries with Real symmetry $\JR$. The instability scenarios of Proposition~\ref{prop-scenarios} combined with Kramers degeneracy allow to define global homotopy invariants for $\GUM(\Kk,\JF,\JR)$. 

\begin{theo}
\label{the-RealInvariants} Let $T\in\GM\UM(\Kk,\JF,\JR)$ on a Real Krein space of kind $(\etaR,\etaFR)$.

\begin{enumerate}[\rm (i)]

\item For $(\etaR,\etaFR)=(1,1)$, $\Sig(T)\in\ZM$ and there is a secondary invariant defined by
$$
\Sec(T)
\;=\;
\Sig(1,T)\,\mbox{\rm mod}\,2
\;\in\;\ZM_2\;.
$$

\item For $(\etaR,\etaFR)=(-1,-1)$, $\Sig(T)=0$ and a $\ZM_2$-invariant is defined by
\begin{equation}
\label{eq-Sig2Def}
\Sig_2(T)
\,=\,
\Big(
\tfrac{1}{2}\sum_{\lambda\in\sigma(T)\cap\SM^1}
\bigl(\nu_+(\lambda,T)+\nu_-(\lambda,T)\bigl)\Big)
\mbox{\rm mod }2
\,=\,
\Big(
\tfrac{1}{2}\,\dim(\Ee_{\SM^1})\Big)
\mbox{\rm mod }2
\;.
\end{equation}

\item For $(\etaR,\etaFR)=(-1,1)$, $\Sig(T)\in 2\,\ZM$.

\item For $(\etaR,\etaFR)=(1,-1)$, $\Sig(T)=0$.

\end{enumerate}
\end{theo}

\noindent {\bf Proof.} (i) As $\nu_\pm(\lambda,T)=\nu_{\pm}(\overline{\lambda},T)$ from the upper to the lower part of the unit circle. One cannot conclude that the signature $\Sig(T)$ is even, however, because it may well happen that at $\lambda=\pm 1$ one has odd inertia and the symmetry says nothing about these points. In particular, one may have either $1$ or $-1$ or both as simple eigenvalues, and neither of them can be lifted by any generic collision scenario. This shows that the secondary invariant is well-defined. Indeed, if $\Sig(T)$ is odd, then exactly one of $\Sig(1,T)$ and $\Sig(-1,T)$ is odd. On the other hand, if $\Sig(T)$ is even, then both $\Sig(1,T)$ and $\Sig(-1,T)$ are either even or odd. (ii) and (iv) Here $\nu_\pm(\lambda,T)=\nu_{\mp}(\overline{\lambda},T)$ by Proposition~\ref{prop-JunitaryTRI}(v). This immediately implies that $\Sig(T)=0$. For (iv) this allows to move all eigenvalues off the unit circle via TB or PD. The finite dimensional perturbations accomplishing this can be constructed in a similar manner as in Section~\ref{sec-InvJherm} for $\JF$-hermitian operators. For (ii), namely $\etaR=-1$, one furthermore has Kramers degeneracy for any real eigenvalue which implies as in the proof of Proposition~\ref{prop-scenarios} that there is no TB and no PD. Thus only the QKC is allowed, which always involves $4$ eigenvalues. This allows either to remove all eigenvalues from $\SM^1$, or to remove all but $2$. These two cases are distinguished by the the invariant $\Sig_2(T)$. (iii) For any $T\in \GUM(\Kk,\JF,\JR)$ one has again $\nu_\pm(\lambda)=\nu_{\pm}(\overline{\lambda})$. In particular, each such pair gives an even contribution (in $2\,\ZM$) to the global signature $\Sig(T)$. The reflection symmetry says little about $\lambda=\pm 1$, but as $\etaR=-1$ these eigenvalues do have Kramers degeneracy, namely by Proposition~\ref{prop-JunitaryTRI}(iv) the generalized eigenspaces of the eigenvalues $\pm 1$ are always of even dimension. As the quadratic from $\JF$ is always non-degenerate, this implies that the Krein signatures $\Sig(\pm 1,T)$ are also always even. In conclusion, the global signature $\Sig(T)$ is always even. 
\hfill $\Box$

\vspace{.2cm}

\noindent{\bf Remark} 
In the case $(\etaR,\etaFR)=(1,1)$ the eigenvalues $\pm 1$ with odd algebraic multiplicity cannot be removed by a perturbation (which may and generically does lift the degeneracy though).  Let us give a finite dimensional example illustrating the appearance of such eigenvalues and their stability. The family of matrices
\begin{equation}
\label{eq-finex}
T_t
\;=\;
\lambda\;
\begin{pmatrix}
\sigma \cosh(t) & \sigma' \sinh(t) \\
-\sigma' \sinh(t)  & -\sigma\cosh(t) 
\end{pmatrix}
\;,
\qquad
\sigma,\sigma'\in\{-1,1\}\;,\;\;\;t\in\RM\;,
\end{equation}
lies in $\mbox{\rm O}(1,1)$ and has eigenvalues $1$ and $-1$. One checks that the Krein signatures are $\Sig(\pm 1)=\pm\sigma$ for all $t$. The global signature $\Sig(T_t)=0$ vanishes,  but for $\mbox{\rm O}(N_+,N_-)$ with $N_++N_-$ odd and in the infinite dimensional situation, one may only have one eigenvalue at $1$ and $-1$ and this could then lead to odd values of the signature. In conclusion,  the signature is a non-trivial invariant for $\GUM(\Kk,\JF,\JR)$ which can take any value in $\ZM$. One may believe that both values of $\sigma$ in \eqref{eq-finex} lead to two non-homotopic realization in the component with $\Sig(T)=0$. Indeed, within $\mbox{\rm O}(1,1)$ no homotopy can be found, but in larger matrices these configurations can be path connected by a succession of a MPD and MTB. Let us stress once again that we do not claim that there is always a homotopy between operators with same global signature and secondary invariant. 
\hfill $\diamond$

\vspace{.2cm}

\noindent{\bf Remark} 
Let us point out that there is a plus sign in the definition \eqref{eq-Sig2Def} of the $\ZM_2$-signature, so that strictly speaking the terminology {\it signature} is not adequate. 
\hfill $\diamond$

\vspace{-.1cm}

\subsection{$J$-hermitian $\RM$-Fredholm operators with Real symmetries}
\label{sec-InvJherm}

Definition~\ref{def-RealKreinColl} of bifurcation scenarios and the arguments in Proposition~\ref{prop-scenarios} transpose from paths of $J$-unitaries to paths of $J$-hermitians via the Cayley transform and \eqref{eq-InertiaCaley} (alternatively the proofs can be repeated). The Cayley transforms $C^{-1}_{z,1}$ and $C^{-1}_{z,-1}$ map the unit circle onto $\overline{\RM}$. The QKC are still generic, but originate from two points $\pm\lambda$ on the real axis. The special points become  $C^{-1}_{z,1}(-1)=C^{-1}_{z,-1}(1)=0$ and $C^{-1}_{z,1}(1)=C^{-1}_{z,-1}(-1)=\infty$. As the $J$-hermitians are bounded by definition, there are only the TB for kind $(1,-1)$ and the MTD for kind $(1,1)$, both through $0$. Afterwards the global invariants for operators in $\FM\HM(\Kk,\JF,\JR)$ can be defined as in Theorem~\ref{the-RealInvariants}, but there is no secondary invariant in the case $(\etaR,\etaFR)=(1,1)$. Again the proofs can be repeated. The following result is a considerable strengthening because it also determines the connected components of $\FM\HM(\Kk,\JF,\JR)$.

\vspace{-.1cm}

\begin{theo}
\label{theo-connectedReal}
Let $(\Kk,\JF,\JR)$ be a Real Krein space of kind $(\etaR,\etaFR)$. For $\tau=1$, the eigenspaces of $\JF$ are supposed to be of same dimension and for the matrix entries $S_\pm$ in \eqref{eq-Schoice} are supposed to be $S_\pm=\one$ for $\eta=1$ and $S_\pm=s$ for $\eta=-1$.

\vspace{-.1cm}

\begin{enumerate}[\rm (i)]

\item For $(\etaR,\etaFR)=(1,1)$, the map $H\in\FM\HM(\Kk,\JF,\JR)\mapsto\Sig(H)\in\ZM$ is a homotopy invariant labelling the connected components.

\item For $(\etaR,\etaFR)=(-1,-1)$, the map $H\in\FM\HM(\Kk,\JF,\JR)\mapsto\Sig_2(H)\!=\!\frac{1}{2}\dim(\Ee_\RM)\,\mbox{\rm mod}\,2\,\in\ZM_2$ is a homotopy invariant labelling the two connected components.

\item For $(\etaR,\etaFR)=(-1,1)$, the map $H\in\FM\HM(\Kk,\JF,\JR)\mapsto\Sig(H)\in 2\,\ZM$ is a homotopy invariant labelling the connected components.

\item For $(\etaR,\etaFR)=(1,-1)$, the set $\FM\HM(\Kk,\JF,\JR)$ is connected.

\end{enumerate}

\end{theo}

The facts that $\Sig$ and $\Sig_2$ are homotopy invariants and that $\Sig$ is even for $(\etaR,\etaFR)=(-1,1)$ follow from the arguments before the statement of the theorem. Hence let us focus on the connectedness statements in the $4$ cases. This will be done by implementing the Real symmetries in the homotopies of proof of Theorem~\ref{theo-Homotopy}. While this is lengthy, the only supplementary ingredient is contained in the following preparation.

\begin{lemma}
\label{lem-unitaryfactorize}
Let $v$ be a unitary on a Hilbert space with real structure, and $s$ be a real operator satisfying $s^*=s^{-1}=-s$.

\begin{enumerate}[\rm (i)]

\item Let $v^t=v$ be symmetric. Then there exists a path $t\in[0,1]\mapsto v_t$ of symmetric unitaries from $v_0=\one$ to $v_1=v$. Furthermore,  there exists a unitary $w$ such that $v=w^tw$. If $v$ is gapped (namely with spectrum not given by all of $\SM^1$), then the path can be chosen to lie in the set of gapped symmetric unitaries. On this space, the map $v\mapsto w$ can be chosen to be continuous.

\item Let $v$ be odd symmetric, namely $s^*v^ts=v$. Then there exists a path $t\in[0,1]\mapsto v_t$ of odd symmetric unitaries from $v_0=s$ to $v_1=v$.  Furthermore, there exists a unitary $w$ such that $u=s^*w^tsw$. If $v$ is gapped, then the path can be chosen to lie in the set of gapped odd symmetric unitaries. On this space, the map $v\mapsto w$ can be chosen to be continuous.

\item Let $v=\overline{v}$ be real with $1\not\in\sigma(v)$. Then there exists a path $t\in[0,1]\mapsto v_t$ of real unitaries with $1\not\in\sigma(v_t)$ from $v_0=-\one$ to $v_1=v$.

\item Let $v=s^*\overline{v} s$ be quaternionic with $1\not\in\sigma(v)$. Then there exists a path $t\in[0,1]\mapsto v_t$ of quaternionic unitaries with $1\not\in\sigma(v_t)$ from $v_0=-\one$ to $v_1=v$.

\end{enumerate} 
\end{lemma}

\noindent {\bf Proof.} (i) Let $v=e^{\imath h}$ for some selfadjoint operator $h$ (by choosing some branch of the logarithm and using spectral calculus) which then satisfies $h^t=h$. Then a path is given by $t\in[0,1]\mapsto v_t=e^{\imath th}$. Set $w=e^{\frac{\imath}{2}h}$ this satisfies $v=w^tw$. For a gapped unitary, $h=-\imath\log(v)$ for an adequate branch of the logarithm so that the dependence on $v$ is continuous. For (ii) let again $v=e^{\imath h}$, then $s^*h^ts=h$. A path is still $v_t=e^{\imath th}$, and the factorization holds with $w=se^{\frac{\imath}{2}h}$. (iii) and (iv) follow by similar arguments, or just using the spectral theorem.
\hfill $\Box$

\vspace{.2cm}

Now let us look at the homotopies in the proof of Theorem~\ref{theo-Homotopy}, and start with the equivalent of Lemma~\ref{lem-Homotopy1}, and then go on by adapting Lemma~\ref{lem-Homotopy2}.

\begin{lemma}
\label{lem-Homotopy1Sym}
For any kind $(\etaR,\etaFR)$, $H\in\FM\HM(\Kk,\JF,\JR)$ is homotopic to  some $H'\in \FM\HM(\Kk,\JF,\JR)$ with $\sigma(H')\subset \{-\imath,0,\imath\}$.
\end{lemma}

\noindent {\bf Proof.}
In fact, the homotopy defined in \eqref{eq-Ht} satisfies $\JR^*\,\overline{H_t}\,\JR=-H_t$ because $H$ has this property and because $\JR^*\,\overline{P_\pm}\,\JR=-P_\pm$ by Proposition~\ref{prop-JunitaryTRI}(ii). 
\hfill $\Box$

\vspace{.1cm}

\begin{lemma}
\label{lem-Homotopy2Sym}
For any kind $(\etaR,\etaFR)\not =(-1,-1)$, $H\in\FM\HM(\Kk,\JF,\JR)$ is homotopic to  some $H''\in \FM\HM(\Kk,\JF,\JR)$ with $\sigma(H'')\subset \{-\imath,0,\imath\}$ such that $0$ is a definite eigenvalue of $H''$. For kind $(\etaR,\etaFR) =(-1,-1)$, the homotopy to $H''\in \FM\HM(\Kk,\JF,\JR)$ with $\sigma(H'')\subset \{-\imath,0,\imath\}$ can be chosen such that either $H''$ has trivial kernel or the inertia of the kernel is $(1,1)$. 
\end{lemma}

\noindent {\bf Proof.} 
Here we look at the perturbations lifting the degeneracies in the proof of Lemma~\ref{lem-Homotopy2}. Again we start from $H=H'$ as given in Lemma~\ref{lem-Homotopy1Sym}. First of all, by Proposition~\ref{prop-JunitaryTRI}(ii) one has $\JR\,\overline{\Ee_{0}(H)}=\Ee_{0}(H)$. Therefore the frame $\Psi$ for $\Ee_{0}(H)$ satisfies $\Psi=\JR\,\overline{\Psi}\, u$ for some unitary matrix $u$. Iterating one gets $\Psi=\etaR \Psi \overline{u}u$ so that $u^t=\etaR u$. For $\etaR=1$, it follows from Lemma~\ref{lem-unitaryfactorize} that there is exists a unitary $v$ such that $u=v^tv$. For $\etaR=-1$, Lemma~\ref{lem-Kramers} shows that $\Ee_{0}(H)$ and thus also $\Psi$ is even dimensional, and thus by Lemma~\ref{lem-unitaryfactorize} there exists a unitary $v$ such that $u=v^t sv$ where  $s=\binom{0 \; -\one}{\one \;\;\; 0}$ is an even-dimensional real orthogonal matrix satisfying $s^2=-\one$. Replacing $\Psi$ by the frame $\Psi v$, one has 
$$
\JR\,\overline{\Psi}\,=\,\Psi\;\;\mbox{\rm for }\;\etaR=1\,, 
\qquad
\JR\,\overline{\Psi}\,=\,-\Psi s\;\;\mbox{\rm for }\;\etaR=-1
\;.
$$

Let us first focus on $\etaR=1$. Then $j_\Psi=\Psi^* \JF\Psi=j_\Psi^*$ is non-degenerate and satisfies
\begin{equation}
\label{eq-JPsiSym}
j_\Psi
\;=\;
\etaFR\;\Psi^*\JR^* \JF\JR\,\Psi
\;=\;
\etaFR\;\overline{j_\Psi}
\;.
\end{equation}
For $\etaFR=1$, the matrix $j_\Psi$ is real symmetric and can hence be diagonalized by a real orthogonal matrix $o$. Replacing $\Psi$ by $\Psi o$, one then has diagonal $j_\Psi$ and still the property $\JR\,\overline{\Psi}=\Psi$. Now one can define $\JF_\Psi=j_\Psi|j_\Psi|^{-1}$ as in \eqref{eq-defJPsi} and consider the perturbation $H_t$ given by \eqref{eq-Ht'bis}. This is still $\JF$-hermitian and the perturbation removes $2n_-$ eigenvalues from the kernel by the argument in Lemma~\ref{lem-Homotopy2} provided that $V$ is chosen $J_\Psi$-hermitian. Moreover, one has to check that $\JR^*\,\overline{H_t}\,\JR=-H_t$, or equivalently $\JR^*\,\overline{\Psi n_\Psi V n_\Psi^{-1}\Psi^* P_0}\,\JR=-\Psi n_\Psi V n_\Psi^{-1}\Psi^* P_0$. This is guaranteed if $V$ purely imaginary because $n_\Psi$ is real and $\JR^*\,\overline{P_0}\,\JR=P_0$ by  Proposition~\ref{prop-JunitaryTRI}(ii). Thus choosing $V$ as in \eqref{eq-V'choice} one has a  homotopy lying in $\FM\HM(\Kk,\JF,\JR)$ for the case $(\etaR,\etaFR)=(1,1)$ which lifts the degeneracy of the kernel so that it is $J$-definite as required. Hence $H''=H_1$ has all the desired properties.

\vspace{.1cm}

Next let us come to the case $(\etaR,\etaFR)=(1,-1)$. As $j_\Psi$ is now imaginary and antisymmetric, one can choose an orthogonal $o$ such that $o j_\Psi o^t=\imath\binom{0 \; -a}{a\;\;\; 0}$ with a diagonal matrix $a>0$. In particular, one necessarily has $n_+=n_-$. Again replacing $\Psi$ by $\Psi o$, one thus has $j_\Psi=\imath\binom{0 \; -a}{a\;\;\; 0}$ as well as $\JR\,\overline{\Psi}=\Psi$. (Note that $\Psi$ has different properties as in the proof of  Lemma~\ref{lem-Homotopy2}.) Hence $J_\Psi=j_\Psi|j_\Psi|^{-1}=\imath\binom{0 \; -\one}{\one\;\;\; 0}$ The perturbation is still of the form \eqref{eq-Ht'}, but now with
\begin{equation}
\label{eq-HtPerturb1-1}
V\;=\;
\begin{pmatrix} \imath\,\one & 0 \\ 0 & -\imath\,\one \end{pmatrix}
\;,
\end{equation}
where each block entry is of size $n_-$. Indeed, this satisfies $V^*J_\Psi=J_\Psi V$ and $\overline{V}=-V$ so that $H_t\in \FM\HM(\Kk,\JF,\JR)$ for the case $(\etaR,\etaFR)=(1,-1)$. Moreover, $\sigma(V)=\{\pm\imath\}$ so that $H''=H_1$ has the desired properties by the spectral argument in the proof of Lemma~\ref{lem-Homotopy2}.

\vspace{.1cm}

Now let $\etaR=-1$. Then \eqref{eq-JPsiSym} is modified to  $j_\Psi=(j_\Psi)^* =\etaFR\,s^*\overline{j_\Psi}s$. For $\etaFR=1$, this means that the matrix $j_\Psi$ is a quaterionic selfadjoint matrix and can hence be diagonalized by a quaternionic unitary $u$, namely one satisfying $s^*\overline{u}s=u$. In particular, both inertia $(n_+,n_-)$ are necessarily even (as the multiplicities of eigenvalues of a selfadjoint quaternionic matrix). Now replacing $\Psi$ by $\Psi u$, one has diagonal $\JF_\Psi$, and the property $\JR\,\overline{\Psi}=-\Psi s$ still holds. More precisely, $\Psi$ can be chosen such that $\JF_\Psi=\diag(\JF'_\Psi,\JF'_\Psi)$ with $\JF'_\Psi=\diag(\one_{n'},\one_{n'_0},-\one_{n'})$ where $n'=\frac{1}{2}n_-$, and $n'_0=\frac{1}{2}(n_+-n_-)$. To guarantee that the path $H_t$ of \eqref{eq-Ht'bis} lies in $\FM\HM(\Kk,\JF,\JR)$ for $(\etaR,\etaFR)=(-1,1)$ one now needs $V^*J_\Psi=J_\Psi V$ and $s^*\overline{V}s=-V$ by the argument above. This holds for $V=\diag(v,v)$ with $v$ given by the matrix in \eqref{eq-V'choice} so that $v^*J'_\Psi=J'_\Psi v$ and $\overline{v}=-v$. Then $H''=H_1$ has a $\JF$-definite kernel of dimension $n_+-n_-$.

\vspace{.1cm}

For the remaining case $(\etaR,\etaFR)=(-1,-1)$, one has $j_\Psi=-s^*\overline{j_\Psi}s$ which implies that the positive and negative spectral subspaces of $j_\Psi$ are mapped onto each other. Hence $n_+=n_-$. Furthermore, $j_\Psi$ can be diagonalized $u^*j_\Psi u=\diag(j,-j)$ with a unitary satisfying $u=s^*\overline{u}s$ and diagonal $j>0$. Let us replace $\Psi u$ by $\Psi $, then one has $j_\Psi=\diag(j,-j)$ and still $\JR\,\overline{\Psi}=-\Psi s$. Hence $\JF_\Psi=\diag(\one_{n_+},-\one_{n_+})$. We need to find $V$ such that $V^*\JF_\Psi=\JF_\Psi V$ and $s^*\overline{V}s=-V$, and $\sigma(V)\subset \{0,\pm\imath\}$. Thus we set $V=\binom{0 \; v}{v\; 0}$ with, for $n_+$ even or odd respectively
$$
v\;=\;
\begin{pmatrix} 0 & \one \\ 
\one & 0 \end{pmatrix}
\;,
\qquad
v
\;=\;
\begin{pmatrix} 0 & 0 & -\one \\ 0 & 0 & 0 \\ \one & 0 & 0 \end{pmatrix}
\;,
$$
where the $\one $ is of size $2[\frac{n_+}{2}]$. As $\overline{v}=v$, one can check all the above properties. Again $H_t$ is in $\FM\HM(\Kk,\JF,\JR)$ and  $H''=H_1$ is as claimed.
\hfill $\Box$

\vspace{.2cm}

\noindent {\bf Proof} of Theorem~\ref{theo-connectedReal}. Now remains to examine the last step in the proof of Theorem~\ref{theo-Homotopy}. We start from the special forms $H=H''$ as given in Lemma~\ref{lem-Homotopy2Sym}. Unless $(\eta,\tau)=(-1,-1)$ and the kernel $\Ee_0(H)$ is of dimension $2$, the projections $P_\pm$ are $J$-Lagrangian by Lemma~\ref{lem-Lag}. For $(\eta,\tau)=(-1,-1)$ and a two-dimensional kernel $\Ee_0(H)$, one can use Lemma~\ref{lem-transformation} and concentrate on the $J_\Phi$-hermitian $H_\Phi$ which has a trivial kernel. In conclusion, in all cases $P_\pm$ are $\JF$-Lagrangian. We now use again the representation \eqref{eq-PRep} of $P_\pm$ in terms of frames $\Phi_\pm$.  Due to the supplementary symmetry $\JR^*\overline{P_\pm}\JR=P_\pm$ following from Proposition~\ref{prop-JunitaryTRI}(ii), the frames now satisfy $\JR\,\overline{\Phi_\pm}=\Phi_\pm w_\pm$ for  adequate unitaries $w_\pm$. Starting from \eqref{eq-LagU}, we use again the particular representation \eqref{eq-Jchoice} for the fundamental symmetry $\JF$. 

\vspace{.1cm}

(i) Due to our hypothesis, $S_\pm=\one$ so that writing out $\JR\,\overline{\Phi_\pm}=\Phi_\pm w_\pm$  then shows that the unitaries $u_\pm$ in \eqref{eq-LagU} are real. The Fredholm property implies that the real unitary $v=u_-^*u_+$ does not have $1$ in the spectrum. Let $v_t$ be the path of real unitaries with $1\not\in\sigma(v_t)$ as given in Lemma~\ref{lem-unitaryfactorize}(iii). Set $u_{+,t}=u_+$ and $u_{-,t}=u_+u_t^*$. Then $u_{-,t}^*u_{+,t}-\one=u_-^*u_+-\one$ is invertible so that the  Lagrangian planes $\Phi_{\pm,t}$ constructed from $u_{\pm,t}$ via \eqref{eq-LagU} span a Fredholm pair. The projections $P_{\pm,t}$ for the projections constructed from $\Phi_{\pm,t}$ by \eqref{eq-PRep} satisfy $\JR^*\overline{P_{\pm,t}}\JR=P_{\pm,t}$. This connects $P_\pm$ as given in Lemma~\ref{lem-Homotopy2Sym} to orthogonal projections $P_{\pm,1}$ through a path respecting all symmetries and the Fredholm property.  Writing out the final projections given by \eqref{eq-PRep} explicitly yields
\begin{equation}
\label{eq-Ppmrep}
P_{\pm,1}
\;=\;
\frac{1}{2}
\begin{pmatrix}
\one & \pm u_+ \\ \pm u_+^* & \one
\end{pmatrix}
\;,
\end{equation}
with a real unitary $u_+$, which in turn can be homotopically deformed into $\one$. In conclusion, we have shown that all operators from $\FM\HM(\Kk,\JF,\JR)$ with same signature can be deformed into one model operator $H'''=\imath(P_{+,1}-P_{-,1})$ where $P_{\pm,1}$ are of the form \eqref{eq-Ppmrep}, but now with $u_+=\one$.

\vspace{.1cm}

(iii) is dealt with in a similar manner by invoking (iv) of Lemma~\ref{lem-unitaryfactorize}.

\vspace{.1cm}

Now we come to the case $\etaFR=-1$. Writing out $\JR\,\overline{\Phi_\pm}=\Phi_\pm w_\pm$  then shows that the unitaries $u_\pm$ in \eqref{eq-LagU} satisfy $(u_\pm)^t=\etaR\, u_\pm$. By Lemma~\ref{lem-unitaryfactorize}(i) this implies that there exist unitaries $v_\pm$ such that $u_\pm=v_\pm^tv_\pm$ for $\etaR=1$, and $u_\pm=v_\pm^tsv_\pm$ for $\etaR=-1$.  First the case $(\etaR,\etaFR)=(1,-1)$ will be considered. The perturbation \eqref{eq-HtPerturb1-1} has completely removed the kernel in this case. The Fredholm condition can be rewritten as the invertibility of $u_-^*u_+-\one=v_+^*(v_+v_-^*(v_+v_-^*)^t-\one)v_+$. Therefore the symmetric unitary $v=v_+v_-^*(v_+v_-^*)^t$ is gapped (because $1$ is not in its spectrum). By Lemma~\ref{lem-unitaryfactorize}(i), we can choose a path $t\in[0,1]\mapsto v_t$ of gapped symmetric unitaries such that $v_0=v$ and $v_1=-\one$. Again by Lemma~\ref{lem-unitaryfactorize} this provides a path $t\in[0,1]\mapsto w_t$ of gapped unitaries such that $v_t=w_t(w_t)^t$. Moreover, a further homotopy in the proof of Lemma~\ref{lem-unitaryfactorize} (of the selfadjoint operator $h$)  allows to choose $w_1=\imath\one$. One has $w_0(w_0)^t=v=v_+v_-^*(v_+v_-^*)^t$, but cannot conclude $w_0=v_+v_-^*$. However, $(v_+v_-^*)^*w_0=\overline{(v_+v_-^*)^*w_0}$ is orthogonal (a real unitary) and therefore by Lemma~\ref{lem-unitaryfactorize} exists a path $t\in[-1,0]\mapsto o_t$ of orthogonals such that $o_{-1}=\one$ and $o_{0}=(v_+v_-^*)^*w_0$. Setting $w_t=v_+v_-^*o_t$ for $t\in[-1,0]$, concatenation yields a path $t\in[-1,1]\mapsto w_t$ from $w_{-1}=v_+v_-^*$ to $w_1=\imath\one$. Finally let us set $u_{+,t}=u_+$ and $u_{-,t}=w_t^*u_+\overline{w_t}$. Both are symmetric unitaries and by construction the invertibility of $u_{-,t}^*u_{+,t}-\one$ is guaranteed for all $t\in[-1,1]$. Furthermore, we have $u_{-,1}=-u_+$ and
$$
u_{-,-1}
\;=\;
w_{-1}^*u_+\overline{w_{-1}}
\;=\;
(v_+v_-^*)^*
(v_+v_+^t) \overline{v_+v_-^*}
\;=\;
v_-v_-^t
\;=\;u_-
\;,
$$
Now let again $\Phi_{\pm,t}$ be the Lagrangian planes constructed from $u_{\pm,t}$ via \eqref{eq-LagU}. Due to the invertibility of $u_{-,t}^*u_{+,t}-\one$ they span a Fredholm pair, and the symmetry of $u_{\pm,t}$ assures $\JR^*\overline{P_{\pm,t}}\JR=P_{\pm,t}$ for the projections constructed from $\Phi_{\pm,t}$ by \eqref{eq-PRep}.  The final projections are again given by \eqref{eq-Ppmrep}, but now with a  symmetric unitary $u_+$. In the final step the unitary can be homotopically deformed in the space of symmetric unitaries to $\one$, again by invoking Lemma~\ref{lem-unitaryfactorize}. In conclusion, have shown that for $(\etaR,\etaFR)=(1,-1)$ every operator from $\FM\HM(\Kk,\JF,\JR)$ can be deformed into one model operator $H'''=\imath(P_{+,1}-P_{-,1})$ where $P_{\pm,1}$ are of the form \eqref{eq-Ppmrep}, but now with $u_+=\one$. This completes the proof of (iv).

\vspace{.1cm}

(ii) Let us now consider the case $(\etaR,\etaFR)=(-1,-1)$. The argument is similar, so let us only focus on the differences. Then $u_-^*u_+-\one= v_+^*s^*(s^*(v_+v_-^*)s(v_+v_-^*)^t-\one)sv_+$ is invertible so that $v=s^*(v_+v_-^*)s(v_+v_-^*)^t$ is a gapped odd symmetric unitary which can be deformed by a path of odd symmetric gapped unitaries $t\in[0,1]\mapsto v_t$ from $v_0=v$ to $v_1=-\one$ (because the projection valued spectral measure is quaternionic). Lemma~\ref{lem-unitaryfactorize} gives a path $t\in[0,1]\mapsto w_t$ of gapped unitaries such that $v_t=s^*w_ts(w_t)^t$ and $w_1=\imath \one$. Now $s^*w_0s(w_0)^t=s^*(v_+v_-^*)s(v_+v_-^*)^t$ so that $(v_+v_-^*)^*w_0=s^*\overline{(v_+v_-^*)^*w_0}s$ is quaterionic (recall that now $s^2=-\one$). Hence there exists a path $t\in[-1,0]\mapsto q_t$ of quaterionic unitaries  such that $q_{-1}=\one$ and $q_{0}=(v_+v_-^*)^*w_0$. Setting $w_t=v_+v_-^*q_t$ for  $t\in[-1,0]$, concatenation yields a path $t\in[-1,1]\mapsto w_t$ from $w_{-1}=v_+v_-^*$ to $w_1=\imath\one$. Next let us set $u_{+,t}=u_+$ and $u_{-,t}=w_t^*u_+\overline{w_t}$. Both are anti-symmetric unitaries and by construction the invertibility of $u_{-,t}^*u_{+,t}-\one$ is guaranteed for all $t\in[-1,1]$. Furthermore, one checks $u_{-,1}=-u_+$ and $u_{-,-1}=u_-$. The associated projection $P_{\pm,1}$ is again of the form \eqref{eq-Ppmrep}. All these projections can homotopically deformed to the case $u_+=s$. Recall, however, that $\FM\HM(\Kk,\JF,\JR)$ has two components due to the first step of the proof.
\hfill $\Box$

\begin{coro}
\label{coro-AtiyahSinger}
Suppose the same hypothesis as in Theorem~\ref{theo-connectedReal}. The set of $\FM\HM(\Kk,\JF,\JR)$ has the following deformation retracts:

\begin{enumerate}[\rm (i)]

\item the set of real Fredholm operators for $(\etaR,\etaFR)=(1,1)$;

\item the set of anti-symmetric Fredholm operators for $(\etaR,\etaFR)=(-1,-1)$;

\item the set of quaternionic Fredholm operators for $(\etaR,\etaFR)=(-1,1)$;

\item the set of symmetric Fredholm operators for $(\etaR,\etaFR)=(1,-1)$.

\end{enumerate}
These sets are respectively the classifying spaces $\FM_0$, $\FM_2$, $\FM_4$ and $\FM_6$ for Real $K$-theory of Atiyah and Singer \cite{AS}.
\end{coro}

\noindent {\bf Proof.} In the proof of Theorem~\ref{theo-connectedReal} we showed that all operators from $\FM\HM(\Kk,\JF,\JR)$ can be homotopically deformed within $\FM\HM(\Kk,\JF,\JR)$ to
$$
H\;=\;\JF\,H^*\,\JF\;=\;-H^*\;=\;-\,\JR^*\,\overline{H}\,\JR
\;.
$$
The first two equations show that $H$ is of the form \eqref{eq-Hspecial}, but with a Fredholm operator $A$ which now satisfies supplementary conditions due to the last identity. Writing these conditions out using the particular representations \eqref{eq-Schoice} leads to the stated symmetries properties of $A$. 
\hfill $\Box$

\section{Unbounded $J$-isometric to $J$-selfadjoint operators}
\label{sec-unbounded}

In this section, we briefly analyze unbounded operators on a Krein space which satisfy a Fredholm property. Real symmetries are not dealt with, as the analysis of Section~\ref{sec-realStruc} transposes directly.

\begin{defini}[\cite{Bog}]  Let $(\Kk,J)$ be a Krein space.
\begin{enumerate}[{\rm (i)}]

\item A closed densely defined operator $T$ with domain $\Dd(T)\subset\Kk$ is called $J$-isometric if for all $\phi,\psi\in\Dd(T)$ one has $\phi^*J\psi=(T\phi)^*J(T\psi)$.

\item An  operator $A$ with domain $\Dd(A)\subset\Kk$ is called $J$-symmetric if for all $\phi,\psi\in\Dd(A)$ one has $\phi^*J(A\psi)=(A\phi)^*J\psi$.

\item A closed densely defined operator $H$ on $\Kk$ is called $J$-selfadjoint if $H=JH^*J$ holds on the domain $\Dd(H)=J\Dd(H^*)$.

\item  The set of $J$-isometric operators is denoted by $\IM(\Kk,J)$.

\item  The set of $J$-selfadjoint operators is denoted by $\SM(\Kk,J)$.

\end{enumerate}
\end{defini}

Let us point to one difference w.r.t. \cite{Bog}, namely here we require $J$-isometric operators to be densely defined and closed. An everywhere defined $J$-isometry is a $J$-unitary,  and an everywhere defined $J$-selfadjoint is $J$-hermitian. Furthermore, every $J$-selfadjoint operator is $J$-symmetric, and an everywhere defined $J$-symmetric operator is $J$-selfadjoint. The sets  $\IM(\Kk,J)$ and $\SM(\Kk,J)$ are endowed with the gap topology \cite{Kat}. Convergence in the gap topology is equivalent to convergence of the resolvent sets and the norm of the resolvents \cite[IV.2.25]{Kat}, and on the subset of bounded operators it coincides with the norm topology. It is now possible to work with Riesz projections and Krein signatures of isolated eigenvalues as before. Hence, one can introduce the set $\GM\IM(\Kk,J)$ of essentially $\SM^1$-gapped $J$-isometries just as in Definition~\ref{def-essS1}, and the set $\GM\SM(\Kk,J)$ of essentially $\RM$-gapped $J$-selfadjoint as in Definition~\ref{def-essR}. For both of these sets the signature can be defined as above. This is a homotopy invariant on $\GM\IM(\Kk,J)$ and $\GM\SM(\Kk,J)$ in the sense that $\Sig$ is constant for continuous paths in $\GM\IM(\Kk,J)$ and $\GM\SM(\Kk,J)$ w.r.t. the gap topology.  The proof of this claim is based on the results of Section IV.3.5 of \cite{Kat} which show that isolated eigenvalues vary continuously in the gap topology. Therefore, all results of Krein collision theory for bounded operators transpose to the unbounded cases. Finally,  following Definition~\ref{def-essS1}, it is also possible to define the set $\FM\IM(\Kk,J)$ of $\SM^1$-Fredholm $J$-isometries and the set $\FM\SM(\Kk,J)$ of $\RM$-Fredholm $J$-selfadjoints. Let us show by an explicit example that there are some surprises for unbounded $J$-selfadjoint operators.

\vspace{.2cm}

\noindent {\bf Example} Section~6.4 of \cite{SB} constructs a loop $t\in[0,2\pi)\mapsto T_t=Te^{\imath tK}$ of $J$-unitary $\SM^1$-Fredholm operators from a given $J$-unitary $T$ and compact $J$-selfadjoint operator $K=JK^*J$, with the property that $\sigma(T_t)=r\,\SM^1\cup r^{-1}\SM^1$ for some $r\in(0,1)$ and all $t$ but for two special values $t=\frac{\pi}{2},\frac{3\pi}{2}$; for these special values the annulus $\{z\in\CM\,:\,r<|z|<r^{-1}\}$ completely fills with point spectrum. Let us now consider the Cayley transform $C_{z,\zeta}^{-1}(T_t)$ of this loop. This is a loop inside the unbouned $J$-selfadjoint operators satisfying the $\RM$-Fredholm property. However, at the above special values, the whole real axis belongs to the point spectrum of these operators. Hence we conclude that $\GM\SM(\Kk,J)$ is a proper subset of $\FM\SM(\Kk,J)$. 
\hfill $\diamond$

\vspace{.2cm}

Next let us point out that the statements of Proposition~\ref{prop-Cayley} can be generalized as follows \cite{Bog}: for $z\in\rho(H)$ of a $J$-selfadjoint operator, $C_{z,\zeta}(H)$ is $J$-unitary; if one only requires $z\not\in\sigma_p(H)$ for a $J$-selfadjoint operator $H$,  it is still possible to define $C_{z,\zeta}(H)$ as $J$-isometric operator. Similarly, if $z\in\rho(T)$ of a $J$-isometric operator $T$, then $C_{z,\zeta}^{-1}(T)$ is $J$-hermitian; and if  $\zeta\not\in\sigma_p(T)$ for a $J$-isometric operator $T$, then $C_{z,\zeta}^{-1}(T)$ is a $J$-selfadjoint operator.  Combining these properties with the homotopy arguments in the proof of Theorem~\ref{theo-connected} one can possibly prove the following

\vspace{.2cm}

\noindent {\bf Conjecture:}  The set $\GM\IM_n(\Kk,J)=\{T\in \GM\IM(\Kk,J)\,:\,\Sig(T)=n\}$ is connected.

\begin{thebibliography}{99}
\bibliographystyle{unsrt}





\bibitem[Ati]{Ati} M.~F.~Atiyah,  {\sl $K$-theory and Reality}, Quart. J. Math. Oxford {\bf 16},  367-386 (1966).

\bibitem[AS]{AS} M.~F.~Atiyah, I.~M.~Singer, {\sl Index theory for skew-adjoint Fredholm operators}, Publ. IHES {\bf 37}, 5-26 (1969).


\bibitem[AI]{AI} T.~Ya.~Azizov, I.~S.~Iokhvidov, {\sl Linear operators in spaces with an indefinite metric}, (John Wiley, 1989).

\bibitem[Bog]{Bog} J.~Bogn\'ar, {\sl Indefinite inner product spaces}, (Springer, Berlin, 1974).











\bibitem[GLR]{GLR} I. Gohberg, P. Lancaster, L. Rodman, {\sl Indefinite linear algebra and applications}, (Birkh\"auser, Basel, 2005).


\bibitem[GS]{GS} J.~Grossmann, H.~Schulz-Baldes, {\sl Index pairings in presence of symmetries with applications to topological insulators}, {\tt arXiv:1503.04834}, to appear in Commun. Math. Phys..








\bibitem[HM]{HM} J. E. Howard, R. S. MacKay, {\sl Linear stability of symplectic maps}, J. Math. Phys. {\bf 28}, 1036-1051 (1987).


\bibitem[Kat]{Kat} T. Kato, {\sl Perturbation Theory for Linear Operators}, (Springer, Berlin, 1966).


\bibitem[Kre]{Kre} M. G. Krein, {\sl Principles of the theory of} $\lambda${\sl -zones of stability of a canonical system of linear differential equations with periodic coefficients}, Memory of A.A.~Andronov, pp. 413-498, Izdat. Akad. Nauk SSSR, Moscow, 1955; English Transl. in: M. G. Krein, {\sl Topics in differential and integral equations and operator theory}, (Birkh\"auser, Boston, 1983).














\bibitem[SB]{SB} H. Schulz-Baldes, {\sl Signature and spectral flow of $J$-unitary $\SM^1$-Fredholm operators}, Integral Equations and Operator Theory {\bf 78}, 323-374 (2014).




\bibitem[SV]{SV} H. Schulz-Baldes, C.~Villegas-Blas, {\sl Invariants for J-unitaries on Real Krein spaces
and the classification of transfer operators}, {\tt arXiv:1306.1816v1}, preprint June, 2013.
 


 
\bibitem[YS]{YS} V. A. Yakubovich, V. M. Starzhinskii, {\sl Linear Differential Equations With Periodic Coefficients}, Volume 1, (Wiley, New York, 1975).


\end{thebibliography}
\end{document}